\documentclass{aa}  

\usepackage{graphicx}
\usepackage{txfonts}
\usepackage{lipsum}
\usepackage{subcaption}         
\usepackage{lscape}             
\usepackage{placeins}            
\usepackage{rotating}
\usepackage{tabularx}
\usepackage[colorlinks,citecolor=blue]{hyperref}
\hypersetup{
    colorlinks = true,
    linkcolor = blue,
    anchorcolor = blue,
    citecolor = blue,
    filecolor = blue,
    urlcolor = blue
    }
\usepackage[normalem]{ulem}
\usepackage{siunitx}              
\usepackage{xcolor}                
\usepackage[utf8]{inputenc}       
\DeclareUnicodeCharacter{02BB}{\textokina}    
\newcommand{\textokina}{\raisebox{0.2ex}{\scalebox{0.8}{`}}}

\begin{document}

\title{Extended $J$-band spectroscopy of red supergiants: parameter identifiability and metallicity robustness}
\titlerunning{Identifiability in extended $J$-band RSG spectroscopy}

   \author{M. A. Urbaneja\inst{1}\corrauth{miguel.urbaneja-perez@uibk.ac.at}        
        \and R. P. Kudritzki\inst{2,3}\email{kud@ifa.hawaii.edu}
}
        
   \institute{Universit\"at Innsbruck, Institut f\"ur Astro- und Teilchenphysik, Technikerstr. 25/8, 6020 Innsbruck, Austria%}
                \and Institute for Astronomy, University of Hawai{\textokina}i~at Manoa, 2680 Woodlawn Drive, Honolulu, HI 96822, USA
                \and LMU M\"unchen, Universit\"atssternwarte, Scheinerstr. 1, 81679 M\"unchen, Germany}

   \date{Received September 30, 20XX}

\abstract
{Red supergiants (RSGs) provide a near-infrared metallicity
diagnostic complementary to optical hot-supergiant spectroscopy.
Their low-resolution extragalactic use requires tests of parameter
identifiability and modelling robustness.}
{We quantify the information content of extended \(J\)-band RSG
spectra at \(R \simeq 2000\) over
\(1.155\)--\(1.340\,\mu\mathrm{m}\) and assess the recovery of
effective temperature, surface gravity, microturbulence, and
metallicity.}
{We evaluate MARCS/TURBOSPECTRUM spectra via an emulator.
A six-node Fisher analysis maps the parameter geometry and
wavelength distribution of information, while controlled mocks
test recovery, continuum placement, and three fixed carbon, nitrogen,
and oxygen (CNO) mixtures.
We analyse 21 SpeX spectra of 20 RSG candidates towards Per~OB1,
using a model-guided continuum treatment, a luminosity-based
gravity prior, and diagonal and regularised correlated-residual
likelihoods.}
{All four parameters yield measurable spectral responses, but
only 8--10\% of the raw metallicity information and 4--7\% of
the gravity information are independent of the other directions.
Extending beyond \(1.22\,\mu\mathrm{m}\) supplies median fractions of 
60\% of the raw
and 66\% of the nuisance-orthogonal metallicity information.
Surface gravity dominates the weakest Fisher mode, though its
prior choice shifts association means by at most 0.03~dex.
Matched mocks recover metallicity with a root-mean-square error
of \(0.04\)~dex. Standard continuum treatment biases
solar-metallicity mocks by about \(-0.23\)~dex, reduced to about
\(-0.01\)~dex by our adopted treatment. Plausible CNO mismatch
adds a zero-point sensitivity of 0.04--0.05~dex, and regularised
covariance changes association means by at most 0.03~dex. The
preferred analysis gives \(\langle[Z]\rangle \simeq -0.10\)~dex,
with a star-to-star dispersion of 0.09~dex.}
{Extended low-resolution \(J\)-band RSG spectra yield recoverable
metallicity and temperature constraints once continuum placement
and weak parameter directions are controlled. Per~OB1 serves as a
validation benchmark, though the absolute scale remains
conditional on the adopted stellar-atmosphere and line-formation
framework.}
 
\keywords{stars: abundances -- stars: fundamental parameters -- supergiants -- infrared: stars -- techniques: spectroscopic -- methods: data analysis}

\maketitle
\nolinenumbers   

\section{Introduction}
\label{sec:introduction}

The chemical composition of young stellar populations traces the
present-day enrichment of star-forming galaxies. Quantitative
spectroscopy of luminous supergiants extends such measurements beyond
the Local Group while retaining the physical insight provided by
individual stars. Optical analyses of blue supergiants have
established this approach as a powerful means of measuring stellar
metallicities and abundance gradients in external galaxies
\citep[e.g.][]{Urbaneja2005,Kudritzki2008,Kudritzki2024,Bresolin2025}.
Red supergiants (RSGs) provide a complementary diagnostic because they
are among the brightest stars at near-infrared wavelengths and
dominate the near-infrared light of young stellar populations
\citep[e.g.][]{Lancon2008,Gazak2013}. Their spectra can therefore be
studied individually \citep[e.g.][]{Patrick2015,Gazak2015} or, at
greater distances, through the integrated light of RSG-dominated young
massive and super star clusters \citep[e.g.][]{Lardo2015,Davies2017}.
The capabilities of James Webb Space Telescope (JWST)
Near-Infrared Spectrograph \citep[NIRSpec,][]{Jakobsen2022} and future 
Extremely Large Telescope instruments, including Multi-Object
Spectrograph \citep[MOSAIC,][]{Hammer2021}, make it increasingly important to establish how reliably such spectra can constrain extragalactic metallicities.

Among the near-infrared windows available for RSG spectroscopy, the $J$-band offers a particularly promising route. The classical diagnostic interval introduced by \citet{Davies2010}, approximately $1.15$--$1.22\,\mu\mathrm{m}$, contains strong atomic features associated principally with Fe, Ti, Si, and Mg, while avoiding some of the stronger molecular contamination found elsewhere in cool-star spectra. These features retain useful abundance sensitivity at resolving powers of only a few thousand \citep{Gazak2014}, making the technique attractive for observations of individual RSGs and unresolved young stellar populations. Previous studies have demonstrated that such spectra can provide metallicity estimates well beyond the distances accessible to conventional high-resolution stellar spectroscopy 
\citep[e.g.][]{Gazak2015,Patrick2017}. A complementary question has received less attention: which parameter combinations are independently constrained at this resolution, and which parts of the spectrum carry metallicity information that remains distinguishable from the responses to other atmospheric parameters?

An acceptable spectral fit does not by itself establish that all
inferred parameters are independently identifiable. At $R \simeq
2000$, individual transitions are blended and the pseudo-continuum is
difficult to define, while changes in effective temperature, surface
gravity, microturbulent velocity, and metallicity can produce
partially similar spectral responses. A strong response to a parameter
therefore does not necessarily imply that the spectrum can distinguish
its effect from those of the other parameters. Continuum distortions
can be projected onto the inferred line strengths, while the assumed
carbon, nitrogen and oxygen (CNO) abundance pattern can modify the molecular background and local pseudo-continuum. The treatment of correlated residuals and limitations of the adopted line-formation framework can introduce further shifts. Consequently, neither an acceptable fit nor small formal uncertainties alone demonstrate robust parameter recovery.

Most foundational RSG $J$-band studies concentrated on the restricted interval near $1.15$--$1.22\,\mu\mathrm{m}$, whereas modern near-infrared spectrographs can provide substantially broader simultaneous coverage. Additional wavelengths are most valuable not merely when they increase the total parameter sensitivity, but when they supply responses that differ from those already present and thereby improve parameter separation. Spectral derivatives and Fisher information provide a direct means of examining this local response geometry, identifying the constrained parameter combinations, and locating the wavelengths that carry independent metallicity information. Because these diagnostics are local and model based, they must be complemented by full inference experiments that test global recoverability and sensitivity to systematic assumptions.

The Perseus~OB1 (Per~OB1) association, which includes the \(h\) and \(\chi\) Persei double cluster, hosts one of the richest
populations of blue and red supergiants in the Galaxy. Most of its massive-star population forms a physically linked structure at
\(d=2.5\pm0.4\)~kpc \citep{deBurgos2020}, while the double cluster and its surrounding halo have a characteristic age of approximately
14~Myr \citep{Currie2010}. Its rich and well-studied massive-star population, together with the availability of complementary spectroscopic constraints, makes it particularly well suited to testing the robustness of extended $J$-band inference under realistic conditions.

In this study, we addressed two related questions. First, what information about effective temperature, surface gravity, microturbulent velocity, and metallicity is present in an extended RSG $J$-band spectrum at $R \simeq 2000$, and where does the independent metallicity information originate? Second, under what assumptions can that information be recovered reliably enough to support the use of RSGs as an independent extragalactic abundance diagnostic? We analysed the interval from approximately $1.155$ to $1.340\,\mu\mathrm{m}$, extending substantially redwards of the classical restricted window. We first used spectral derivatives and Fisher information to identify the locally constrained parameter combinations and map the wavelength distribution of raw and nuisance-orthogonal metallicity information. We then performed full Bayesian analyses of controlled mock observations to test parameter recovery and isolate the effects of continuum normalisation and CNO abundance patterns. Finally, we applied the framework to SpeX spectra of RSG candidates towards Per~OB1 and examined the sensitivity of the inferred metallicity scale to the gravity prior, abundance pattern, and treatment of correlated residuals.

Sect.~\ref{sec:methods} describes the observations and spectral models. Sects.~\ref{sec:fisher} and \ref{sec:mocks} present the information analysis and mock-recovery experiments, respectively, while Sect.~\ref{sec:perob1} gives the application to Per~OB1. We discuss the implications for low-resolution RSG abundance work in Sect.~\ref{sec:discussion} and summarise our conclusions in Sect.~\ref{sec:conclusions}.

\section{Observations and spectral models}
\label{sec:methods}

\subsection{SpeX observations and Per~OB1 sample}
\label{sec:observations}

Our sample comprises 21 spectra of 20 red supergiant candidates towards Per~OB1. We observed the targets on 11 October 2011 with the NASA Infrared Telescope Facility (IRTF) SpeX spectrograph 
\citep[][]{Rayner2003} in its short cross-dispersed (SXD) configuration. This mode records approximately 0.8--2.5~$\mu\mathrm{m}$ simultaneously, although we restricted the present analysis to 
three intervals spanning 1.155--1.340~$\mu\mathrm{m}$, which we refer to as the extended $J$-band region. We used a $0\farcs3$ slit, which provided a nominal resolving power of $R \sim 2000$ with approximately three detector pixels per resolution element. The targets were nodded along the slit for sky subtraction, and A0V telluric standards were observed at similar airmasses and close in time to the science exposures.

We reduced the data with the Spextool pipeline \citep{Cushing2004}, including flat-field correction, wavelength calibration, one-dimensional extraction, and combination of the nodded exposures. We corrected telluric absorption with the A0V standards following \citet{Vacca2003}. Because residual structure can remain in regions of strong atmospheric absorption and around hydrogen lines intrinsic to the standards, we excluded the affected intervals from the fit.

We assessed association membership by cross-matching the targets with the catalogue of \citet{deBurgos2020}. That study defined its
membership criteria and association locus using Gaia Data Release 2
(DR2) astrometry \citep{GaiaCollaboration2018}; we retained the same data release for
the four unclassified targets to ensure a homogeneous comparison. Sixteen stars occur in their Table~A.4: nine 
are confirmed members, six are likely members, and BD+56~724 is a candidate, while none is classified as a non-member. The likely members have formally discrepant Gaia DR2 parallaxes, but de Burgos et al. noted that their astrometry may be affected by the large angular sizes and photocentric variability of red supergiants. BD+56~724 is less secure because both its parallax and proper motion are discrepant, although its radial velocity and apparent magnitude agree with those of the other association red supergiants. Of the four targets without a classification in Table~A.4, HD~13658 and BD+57~647 have Gaia DR2 astrometry compatible with the association locus, BD+58~501 has a compatible proper motion but a comparatively high parallax, and HD~236947 was not assessed by de Burgos et al. We retained the complete sample and flagged BD+56~724, BD+58~501, and HD~236947 as the objects whose association membership remains least secure.

One target, BD+59~372, was observed twice during the night. We retained both spectra as an internal repeatability test but represented the star only once when calculating association-level statistics.

Within this wavelength range, we excluded the SpeX coverage
shortward of \(1.155\,\mu\mathrm{m}\), including the more densely blended $Y$-band region; the systematic CN residual 
near 1.1$\,\mu\mathrm{m}$ discussed in Appendix~\ref{app:perob1} illustrates why extending the inference into this region would require
separate validation. Redward of 1.340$\,\mu\mathrm{m}$, strong atmospheric absorption separates the $J$ and $H$ windows, while
the $H$- and $K$-band spectra introduce a different, more molecule-dominated diagnostic regime. Their inclusion lies beyond the
scope of the present $J$-band analysis. The resulting extended coverage is distinct from the restricted
1.15--1.22$\,\mu\mathrm{m}$ window of the classical $J$-band technique. To preserve comparability with the diagnostic selection of
\citet{Gazak2014}, and because the Mg\,\textsc{i} feature near
1.182$\,\mu\mathrm{m}$ is known to exhibit substantial non-LTE effects in
RSG atmospheres \citep{Bergemann2015}, we assigned zero weight to this feature. We applied the same mask to the Fisher
calculations, mock experiments, and observed spectra.

\subsection{Synthetic spectral grids and MAUI emulator}
\label{sec:models}

We computed the synthetic spectra with TURBOSPECTRUM
\citep{Alvarez1998,Plez2012} from one-dimensional, hydrostatic,
spherical MARCS atmospheres calculated for red supergiants
\citep{Gustafsson2008}, assuming local thermodinamyc equilibrium (LTE) line formation. The MARCS models adopted a stellar mass of $15\,M_\odot$. We varied the effective temperature, $T_{\rm eff}$, surface gravity, $\log g$, synthesis microturbulence, $\xi$, and global metallicity coordinate, $[Z]$, defining
\[
\boldsymbol{\theta}_{\rm atm}
=
\left(T_{\rm eff},\log g,\xi,[Z]\right).
\]
The adopted parameter ranges were
\[
\begin{aligned}
3300 &\leq T_{\rm eff} \leq 4500\,\mathrm{K},\\
-0.5 &\leq \log g \leq 1.0\,\mathrm{dex},\\
0.5 &\leq \xi \leq 6.5\,\mathrm{km\,s^{-1}},\\
-1.0 &\leq [Z] \leq +0.5\,\mathrm{dex}.
\end{aligned}
\]
The grid spacing was 100~K in $T_{\rm eff}$, 0.5~dex in $\log g$,
0.5~$\mathrm{km\,s^{-1}}$ in $\xi$, and 0.25~dex in $[Z]$. The MARCS
structures formed a regular three-dimensional grid in $T_{\rm eff}$,
$\log g$, and $[Z]$ with a fixed atmospheric microturbulence of
$\xi_{\rm atm}=5\,\mathrm{km\,s^{-1}}$, while the spectral synthesis sampled $\xi$ from 0.5 to 6.5~$\mathrm{km\,s^{-1}}$. Appendix~\ref{app:grid} describes this construction and its scope in detail, and Sect.~\ref{sec:discussion} discusses the implications of known non-LTE effects.

The MARCS structures used the solar reference composition of \citet{Grevesse2007}, for 
which \(\epsilon_{\odot}(\mathrm{Fe})=7.45\),  scaled globally with $[Z]$. We applied the same global scaling in the spectral synthesis before imposing the grid-specific CNO offsets, while iron and all elements without an explicit offset followed $[Z]$. The inferred $[Z]$ is therefore the global metallicity coordinate of the adopted grid rather than an independently measured line-by-line iron abundance. This distinction is important when comparing our results with literature $[\mathrm{Fe/H}]$ measurements based on different abundance mixtures, solar references, or spectral diagnostics.

We constructed three otherwise equivalent grids with fixed CNO mixtures:
\[
\begin{aligned}
{\rm solar:}\quad
&([\mathrm{C/Fe}],[\mathrm{N/Fe}],[\mathrm{O/Fe}])
=(0.0,0.0,0.0),\\
{\rm mild:}\quad
&([\mathrm{C/Fe}],[\mathrm{N/Fe}],[\mathrm{O/Fe}])
=(-0.2,+0.5,0.0),\\
{\rm heavy:}\quad
&([\mathrm{C/Fe}],[\mathrm{N/Fe}],[\mathrm{O/Fe}])
=(-0.4,+0.7,-0.1).
\end{aligned}
\]
The solar grid retained the solar carbon isotopic ratio, while both processed grids adopted ${}^{12}\mathrm{C}/{}^{13}\mathrm{C}=10$. The relative CNO offsets and isotopic ratio remained fixed as $[Z]$ varied, and we treated the three mixtures as discrete modelling alternatives rather than as individual CNO measurements or a continuous processing coordinate.

We adopted the mildly processed mixture as the reference composition for Per~OB1. Its carbon depletion and approximately solar oxygen abundance are close to the mean values reported for the association RSGs by \citet{Fanelli2022}, while the overall CNO pattern is broadly consistent with the surface processing predicted near the onset of the RSG phase in rotating solar-metallicity models of approximately \(12\)--\(15\,M_\odot\) \citep{Ekstroem2012}. Its somewhat stronger nitrogen enhancement remains representative of plausible RSG surface processing. The mild grid is therefore a physically motivated reference rather than an exact description of every star, while the heavy grid provides a deliberately stronger test of abundance-pattern sensitivity.

At fixed $T_{\rm eff}$, $\log g$, and $[Z]$, the three CNO grids shared the same scaled-solar MARCS structure, and we introduced the CNO and isotopic changes only during spectral synthesis. Their comparison therefore measures the direct spectral sensitivity to the assumed CNO pattern within a fixed-atmosphere framework; Appendix~\ref{app:cno_grids} discusses the excluded structural feedback.

For each CNO grid, we trained a separate statistical emulator using
the Machine-learning assisted Uncertainty Inference framework
\citep[MAUI,][]{Urbaneja2026}. In brief, a principal-component analysis (PCA) of the model grid provided a reduced spectral basis, and Gaussian-process regression mapped the retained coefficients across \(\boldsymbol{\theta}_{\rm atm}\). The emulator predicted continuum-normalised spectra only within the corresponding parameter domain, and extrapolation was explicitly forbidden. The three CNO mixtures remained separate modelling choices, with no interpolation between them.

\section{Information content of the extended J band}
\label{sec:fisher}

\subsection{Fisher-information framework}
\label{sec:fisher_framework}

We used a Fisher-information analysis \citep{Fisher1922,Fisher1925} to examine the local information carried by the extended $J$-band spectra about the four atmospheric parameters. This framework has long been used for parameter forecasting in cosmology \citep[e.g.][]{Tegmark1997} and has more recently been applied to the information content of stellar spectra \citep[e.g.][]{Ting2017,Sandford2023}. The noise-weighted length of a spectral derivative measures how strongly the spectrum responds to a parameter, whereas the angle between two derivatives measures how readily their effects can be distinguished. Two parameters may therefore produce strong spectral changes but remain difficult to separate if their derivatives have similar shapes.

For a model flux \(f_i\) at pixel \(i\) and parameter vector
\(\boldsymbol{\theta}=(T_{\rm eff},\log g,\xi,[Z])\), we defined the
Jacobian as

\[
J_{i\alpha}=\frac{\partial f_i}{\partial\theta_\alpha}.
\]

Each column \(J_\alpha\) is the spectral derivative with respect to
\(\theta_\alpha\), and therefore describes the local spectral response
to a change in that parameter. We then defined its noise-weighted form as

\[
d_\alpha=C^{-1/2}J_\alpha,
\]

where \(C\) is the diagonal noise covariance adopted for this
diagnostic calculation. The spectral Fisher matrix is the matrix of
inner products between these noise-weighted derivative vectors,

\[
F_{\alpha\beta}^{\rm spec}
=J_\alpha^{\rm T}C^{-1}J_\beta
=d_\alpha^{\rm T}d_\beta .
\]

Its diagonal elements give the squared noise-weighted lengths of the
individual derivatives and therefore quantify the response to each
parameter when the others are held fixed. Its off-diagonal elements
give the corresponding overlaps between pairs of derivatives.

Under a local linear and Gaussian approximation, the inverse spectral
Fisher matrix estimates the parameter covariance matrix,
\[
\Sigma_\theta^{\rm spec}
\simeq \left(F^{\rm spec}\right)^{-1}.
\]
The diagonal elements of \(\Sigma_\theta^{\rm spec}\) give the local
marginal variances of the parameters, that is, their variances after
allowing the other parameters to vary, while the off-diagonal elements
describe their covariances. It is often more informative to consider the 
normalised covariances, that is. the correlation coefficients, 
\[
\rho_{\alpha\beta}
=
\frac{\Sigma_{\alpha\beta}^{\rm spec}}
{\sqrt{\Sigma_{\alpha\alpha}^{\rm spec}
       \Sigma_{\beta\beta}^{\rm spec}}}.
\]
Large absolute values of \(\rho_{\alpha\beta}\) indicate that changes
in two parameters can compensate for one another locally in the
spectrum. Complementarily, the eigenvectors of \(F^{\rm spec}\)
identify constrained combinations of parameters, and the eigenvector
associated with the smallest eigenvalue gives the least distinguishable
spectral direction.

To examine how a physically motivated external gravity information used in the full inference modifies the local information geometry,
we included a Gaussian luminosity-based gravity prior through
\[
F^{\rm post}=F^{\rm spec}+F^{\rm prior},
\]
where \(F^{\rm prior}\) is the contribution from an external constraint on the surface gravity, taken 
to have an uncertainty of \(\sigma_{\log g}=0.20\)~dex. This constraint is derived from the 
stellar luminosity and effective temperature (see Appendix~\ref{app:gravity}). It stabilises 
the fit where the spectrum alone constrains gravity only weakly, and it allows us to keep track 
of how much of the final information comes from the spectrum and how much from the external constraint.

We also quantified how much of the response to parameter $\alpha$ remained unique after accounting for the other three. Projecting its noise-weighted derivative $\boldsymbol{d}_\alpha$ away from the subspace spanned by the nuisance derivatives gives $\boldsymbol{d}_{\alpha,\perp}$, from which we defined
\[
\eta_\alpha =
\frac{
\boldsymbol{d}_{\alpha,\perp}^{\rm T}
\boldsymbol{d}_{\alpha,\perp}
}{
\boldsymbol{d}_{\alpha}^{\rm T}
\boldsymbol{d}_{\alpha}
}.
\]
Values near unity indicate a largely distinctive response, whereas values near zero indicate that the other parameters can reproduce most of it. We calculated $\eta_\alpha$ from the spectral derivatives before adding the gravity prior, so it measures information supplied by the spectrum itself.

We evaluated six representative RSG nodes with $T_{\rm eff}=3600$, 3800, and 4000~K, $\log g=0.0$, and $\xi=3$~km~s$^{-1}$, at both $[Z]=0.0$ and $-0.5$. The calculations used $R=2000$, three pixels per resolution element, and $\mathrm{S/N}=200$ per pixel over the wavelength selection described in Sect.~\ref{sec:observations}. We applied the model-guided continuum treatment described in Appendix~\ref{app:continuum} before constructing the derivatives.
Appendix~\ref{app:fisher} documents its application to the displaced spectra in the Fisher calculation, together with the parameter
scaling, finite-difference validation, projection algebra, and
node-by-node diagnostics. We stress that these inverse Fisher matrices are used here
primarily as local covariance approximations for diagnosing
parameter coupling and the effect of the prior.

\subsection{Parameter identifiability}
\label{sec:parameter_identifiability}

All four parameters produce measurable spectral responses, but most
of these responses are not mutually independent. To compare their
coupled directions without allowing the numerical units of the
parameters to determine the result, we expressed parameter offsets in
units of \(100\)~K, \(0.1\)~dex, \(1\,{\rm km\,s^{-1}}\), and
\(0.1\)~dex for \(T_{\rm eff}\), \(\log g\), \(\xi\), and \([Z]\),
respectively, before calculating the condition numbers and
eigenmodes.

\begin{figure*}
\centering
\includegraphics[width=.86\textwidth]{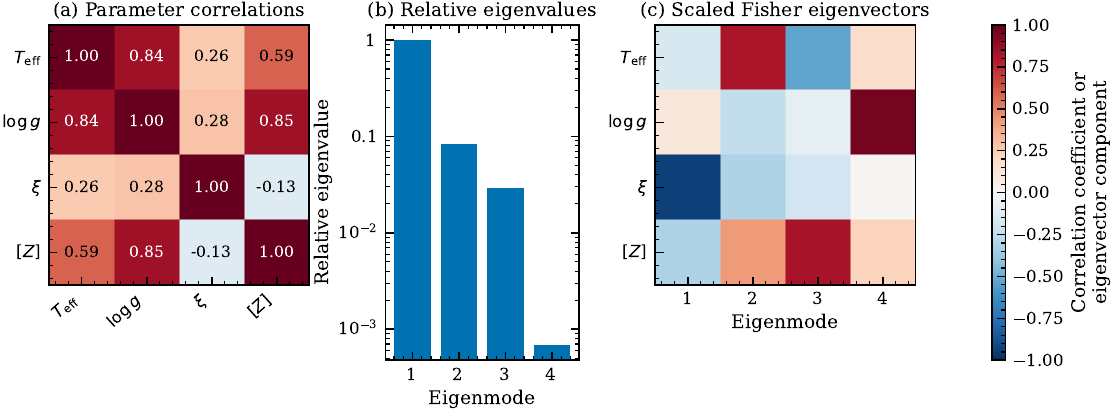}
\caption{Spectral Fisher-information geometry for the representative $T_{\rm eff}=3800\,{\rm K}$, $\log g=0.0$, $\xi=3\,{\rm km\,s^{-1}}$, and $[Z]=0.0$ model at $R=2000$ and ${\rm S/N}=200$ per pixel. Panel~(a) shows the correlations derived from the inverse spectral Fisher matrix, panel~(b) gives the eigenvalues relative to that of the strongest mode, and panel~(c) shows the eigenvector components in the scaled coordinates defined in Appendix~\ref{app:fisher:definitions}. All panels refer to the spectral matrix before addition of the gravity prior, and the eigenmodes are ordered from the strongest to the weakest.}
\label{fig:fisher_correlation}
\end{figure*}

Figure~\ref{fig:fisher_correlation} illustrates the resulting information
geometry for the representative solar-metallicity model at
\(T_{\rm eff}=3800\)~K. The local parameter correlation coefficients derived from the inverse
spectral Fisher matrix, shown in panel~(a), reveal strong positive
correlations between \(\log g\) and \([Z]\) (\(\rho=0.85\)) and
between \(T_{\rm eff}\) and \(\log g\) (\(\rho=0.84\)), together with
a substantial \(T_{\rm eff}\)--\([Z]\) correlation (\(\rho=0.59\)). The relative
eigenvalues in panel~(b) show that the weakest mode carries more than
three orders of magnitude less information than the strongest. Its
scaled eigenvector, shown in panel~(c), is predominantly gravity-like,
demonstrating that surface gravity supplies the dominant contribution
to the least distinguishable parameter combination.

The contrast between the strongest and weakest information directions
can be summarised by the condition number, which is
\(1.46\times10^{3}\) for the spectral-only matrix at this node. This
large value reflects the poorly constrained gravity-dominated
direction. Adding the \(0.20\)~dex gravity prior reduces the condition
number modestly, to \(1.24\times10^{3}\), because the prior adds
information mainly along that weak direction. It therefore 
stabilises the solution, while the inferred gravity remains
prior-informed rather than purely spectroscopic.

The same qualitative structure persists across the six representative
nodes. The independent-information fraction ranges from approximately
\(14\)--\(38\%\) for \(T_{\rm eff}\), \(4\)--\(7\%\) for \(\log g\),
\(31\)--\(35\%\) for \(\xi\), and \(8\)--\(10\%\) for \([Z]\).
The extended \(J\)-band spectrum therefore responds strongly to
metallicity, but most of this response overlaps with changes produced
by the other atmospheric parameters. Surface gravity is the least
independently identifiable parameter and dominates the weakest scaled
Fisher eigenmode at every node, with an absolute gravity coefficient
of \(0.957\)--\(0.964\). Additional calculations at
\(\log g=-0.25\) and \(+0.5\) preserved this structure, showing that
it is not peculiar to the principal choice of \(\log g=0.0\).

Microturbulence is better separated than gravity, although its
influence on line strength and saturation remains coupled to
metallicity. Effective temperature has a more distinctive response
than gravity or metallicity, but its independent fraction varies
across the nodes. The \(T_{\rm eff}\)--\([Z]\) correlation strengthens
towards warmer and more metal-poor models, while the strong
\(\log g\)--\([Z]\) correlation persists throughout the six-node set.

An imposed local offset of \(+100\)~K corresponds to a compensating
metallicity displacement of approximately \(+0.067\) to \(+0.103\)~dex
in the spectral-only geometry and \(+0.059\) to \(+0.095\)~dex after
including the gravity prior. These are local linear responses rather
than empirical corrections, but they show how a coherent
temperature-scale offset can propagate into metallicity. Because the 
idealised calculation assumes a locally exact model
and uncorrelated noise, we use it primarily to diagnose parameter
couplings; Sect.~\ref{sec:mocks} tests quantitatively
how well the corresponding local covariance approximation
reproduces the posterior under matched inference conditions.

\subsection{Distribution of information with wavelength}
\label{sec:fisher_wavelength}

We distinguished the raw, single-parameter information from the nuisance-orthogonal information that remains after removing the spectral directions associated with the other fitted parameters. The former locates wavelengths at which the spectrum responds strongly to a parameter, whereas the latter locates regions that help distinguish that response from competing atmospheric changes. Appendix~\ref{app:fisher:eta} gives the formal definitions and wavelength-resolved diagnostics.

\begin{figure}
\centering
\includegraphics[]{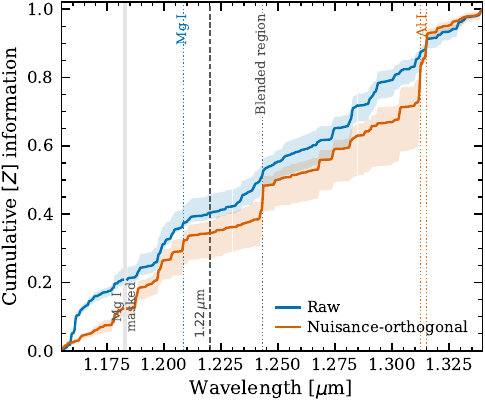}
\caption{Cumulative metallicity information across the extended $J$-band interval. The curves show the median cumulative raw information (blue) and nuisance-orthogonal information (orange) for $[Z]$ across the six atmospheric nodes, while the shaded regions span their full ranges. Each curve is normalised by its total over $1.155$--$1.340\,\mu{\rm m}$; the dashed line marks $1.22\,\mu{\rm m}$, the grey region denotes the excluded Mg~{\sc i} interval, and the dotted lines indicate selected Mg~{\sc i} and Al~{\sc i} features 
and the blended region near $1.243\,\mu{\rm m}$}.
\label{fig:metallicity_information}
\end{figure}

Figure~\ref{fig:metallicity_information} shows the cumulative metallicity information across the analysed interval. By $1.22\,\mu{\rm m}$, approximately the red boundary of the classical restricted $J$-band, the six nodes have accumulated a median of about 40\% of the raw information and 34\% of the nuisance-orthogonal information. The red extension consequently supplies about 60\% and 66\%, respectively, with the node-to-node values given in Table~\ref{tab:fisher_wavelength_results}. The classical \(1.155\)--\(1.220\,\mu\mathrm{m}\) interval occupies
approximately 35\% of the complete wavelength span, so these integrated fractions are close to proportional to the respective
wavelength coverage. This broad correspondence does not imply a uniform information density at the pixel level, because the cumulative curves contain localised contributions from particular blended regions. 

The wavelength-resolved diagnostics in Appendix~\ref{app:fisher:eta} show that the largest nuisance-orthogonal metallicity contributions include blends near $1.243\,\mu{\rm m}$ and the Al~{\sc i} features around $1.312$--$1.315\,\mu{\rm m}$. At $R\simeq2000$, these peaks represent the collective response of blended regions and cannot be assigned unambiguously to individual transitions. Their value lies in having parameter dependences that differ from those of the nuisance directions, rather than simply in producing deep absorption features.

The classical interval therefore contains a substantial metallicity
response, but only about one-third of the information that remains
distinct from the other atmospheric responses. The gain from the red
extension is largely cumulative, reflecting its broader wavelength
coverage, while localised features within it provide additional
responses that help distinguish metallicity from the nuisance
directions. Extending the
wavelength coverage materially improves the separation of metallicity
from the nuisance directions, although the small global value of
\(\eta_{[Z]}\) shows that substantial coupling remains.

\section{Spectral inference and mock recovery}
\label{sec:mocks}

The Fisher diagnostics describe the local response geometry around
each atmospheric node, but they do not test parameter recovery through
the complete non-linear inference or its sensitivity to analysis
choices. We therefore used controlled mock observations to determine
whether metallicity remains recoverable and to quantify the effects of
continuum placement and an incorrect CNO mixture.

\subsection{Forward model and spectral inference}
\label{sec:inference}

At each trial point, the emulator predicted the intrinsic normalised spectrum for the atmospheric parameter vector $\boldsymbol{\theta}_{\rm atm}=(T_{\rm eff},\log g,\xi,[Z])$. We convolved this spectrum with the instrumental line-spread function, transformed it onto the wavelength scale of the analysed spectrum, and resampled it onto the observed pixels. A scale factor $s_R$ allowed modest deviations from the nominal resolving power through $R_{\rm eff}=s_RR_{\rm nom}$, where $R_{\rm nom}=2000$. The deterministic parameters were therefore $(T_{\rm eff},\log g,\xi,[Z],s_R)$, while the CNO mixture was fixed by the selected grid.

The forward model included a trial-dependent, model-guided continuum
correction based on the principle introduced by \citet{Gazak2014}. We identified pseudo-continuum anchors from the current trial 
spectrum and recalculated the correction at every parameter vector; we refer to this implementation as the Gazak-like continuum treatment. For the controlled comparison in Sect.~\ref{sec:continuum_mocks}, we
also implemented the explicit model-guided algorithm given by \citet{Gazak2014PhDT}, which we refer to as the exact-Gazak
treatment. Appendix~\ref{app:continuum} gives the complete algorithm.

For the observed stars, we adopted a Gaussian gravity prior centred on
the luminosity-based estimate for each object, with
\(\sigma_{\log g}=0.20\)~dex; Appendix~\ref{app:gravity} gives its
construction. For the mocks, we adopted a Gaussian prior with the same
width but centred on the injected value, \(\log g=0.0\). In both cases,
the prior supplied external information along the weak
gravity-dominated direction identified in Sect.~\ref{sec:fisher}
without fixing gravity to a single value. We assigned uniform priors
to \(T_{\rm eff}\), \(\xi\), and \([Z]\) within the grid, prohibited
extrapolation, and adopted \(s_R\sim{\cal N}(1,0.15^2)\).

We considered a diagonal Gaussian likelihood, with observational and
emulator variances added at each pixel, and a regularised likelihood
that added a Matérn-3/2 covariance with relative amplitude
\(s_{\rm cov}\) and sampled correlation length \(\ell_{\rm cov}\).
The latter was expressed in pixels and converted locally to a velocity
scale when evaluating the covariance kernel. We sampled one pair of
covariance hyperparameters shared by the three fitted intervals, with 
\[
\log s_{\rm cov}\sim{\cal N}(-1,0.3^2),
\quad
-1.8\leq\log s_{\rm cov}\leq-0.2,
\]
and
\[
\ell_{\rm cov}\sim{\cal N}(2,1^2),
\quad
1\leq\ell_{\rm cov}\leq6\ {\rm pixels}.
\]
Appendix~\ref{app:covariance} gives the likelihood, kernel normalisation, and covariance diagnostics. The diagonal form exposed how continuum or abundance-pattern mismatches were projected onto the atmospheric parameters, whereas the regularised covariance allowed a controlled fraction of coherent residual structure to remain outside the atmospheric model. We used the former for the controlled mocks and compared both prescriptions for the observed spectra.
In all cases, we summarised each marginal posterior by its median and 16th and 84th percentiles. 

\subsection{Mock-observation design}
\label{sec:mock_design}

We constructed the CNO mocks at the six primary Fisher nodes: $T_{\rm eff}=3600$, 3800, and 4000~K, $\log g=0.0$, $\xi=3$~km~s$^{-1}$, and $[Z]=0.0$ and $-0.5$. We degraded the spectra to $R=2000$, sampled them at three pixels per resolution element, and added Gaussian noise corresponding to $\mathrm{S/N}=200$ per pixel. We used the same wavelength windows, Mg~{\sc i} mask, and gravity prior as in the corresponding Fisher calculation.

For the continuum experiment, we generated ten independent noise realisations at each metallicity for the $T_{\rm eff}=3800$~K node. We analysed every realisation with the standard, Gazak-like, and exact-Gazak continuum treatments, using the same noise realisation for all three methods and otherwise matching the inference configuration. The 60 resulting analyses therefore isolated the continuum prescription.

For the CNO experiment, we generated one mock from each of the three mixtures at every physical node, giving 18 spectra, and fitted every mock with all three grids, giving 54 posterior analyses. Matched generating and fitting grids tested numerical and non-linear recoverability, while the off-diagonal combinations measured how a fixed CNO mismatch was redistributed among the inferred parameters. We paired the noise realisations across generating mixtures and retained the diagonal likelihood so that differential shifts were projected directly onto the model parameters rather than absorbed by a correlated-residual component.

\subsection{Continuum normalisation}
\label{sec:continuum_mocks}

Continuum placement is particularly important in line-blanketed RSG
spectra. The standard procedure anchors the pseudo-continuum to the highest points of the noisy observation, whereas both the Gazak-like and exact-Gazak methods select anchors from each trial model and thereby avoid preferentially choosing pixels displaced upwards by noise.

\begin{figure}
\centering
\includegraphics[width=\columnwidth]{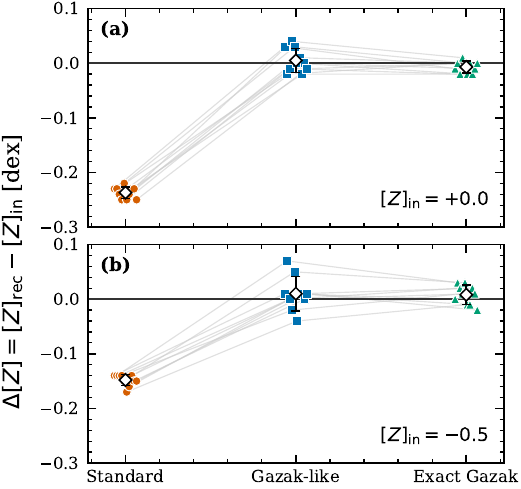}
\caption{Metallicity recovery from ten paired noise realisations using the standard, Gazak-like, and exact-Gazak continuum treatments at $T_{\rm eff}=3800$~K, $\log g=0.0$, $\xi=3$~km~s$^{-1}$, and $\mathrm{S/N}=200$ per pixel. Panels~(a) and (b) show $[Z]_{\rm in}=0.0$ and $-0.5$, respectively, and the horizontal line marks exact recovery.}
\label{fig:continuum_mock}
\end{figure}

Figure~\ref{fig:continuum_mock} shows the resulting metallicity recovery. At $[Z]=0.0$, the standard procedure produced a mean displacement of approximately $-0.23$~dex, while the Gazak-like method reduced it to approximately $-0.01$~dex and the exact-Gazak implementation was likewise essentially unbiased. The same ordering persisted at $[Z]=-0.5$, and the paired noise realisations show that the movement between methods was not caused by different random samples.

Continuum placement is therefore the largest systematic vulnerability isolated by the mock experiments. Selecting anchors from the trial model avoids defining the pseudo-continuum from noise-enhanced data and better preserves the relative depths of metallicity-sensitive blends. The agreement between the two model-guided implementations shows that the improvement follows from their shared principle rather than the details of one implementation. We adopted the Gazak-like continuum treatment for the production analysis.

\subsection{Sensitivity to the CNO mixture}
\label{sec:cno_mocks}
The matched-grid fits recovered the injected scale without a material unexplained bias. Across the 18 matched cases, the mean metallicity offset was \(+0.03\)~dex, the root-mean-square (RMS) error was \(0.04\)~dex,
and the maximum absolute offset was \(0.06\)~dex; all injected metallicities lay within their nominal 68\% intervals. Appendix~\ref{app:mock_diagnostics} gives the exact values, corresponding diagnostics for the other parameters, and qualifications on this consistency check.

\begin{figure*}
\centering
\includegraphics[width=\textwidth]{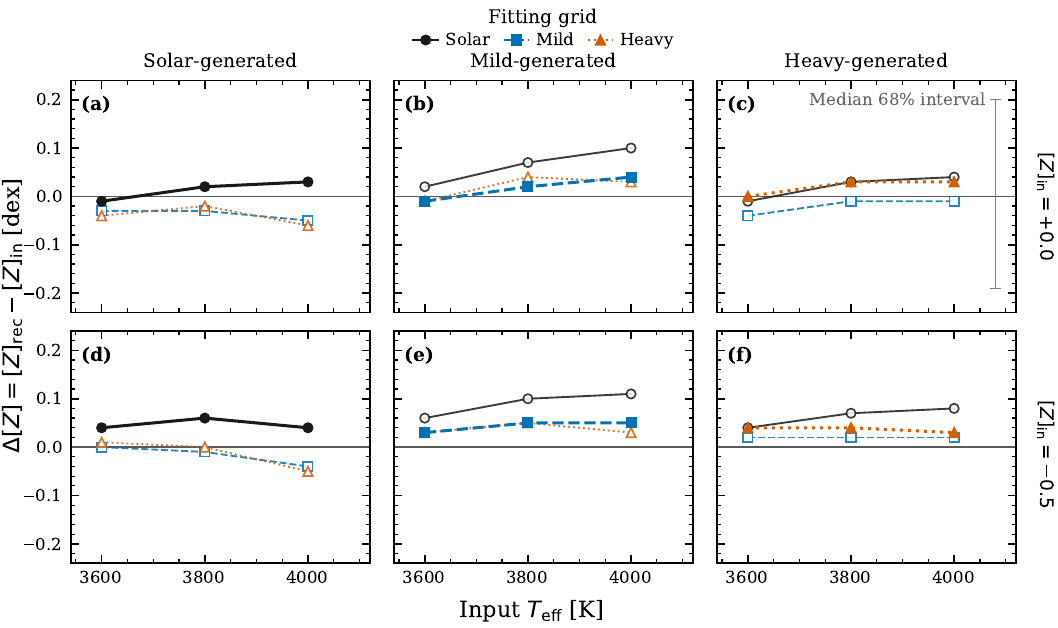}
\caption{Metallicity recovery in the CNO cross-grid experiment. The columns show mocks generated with the solar, mild, and heavy mixtures, and the rows show $[Z]_{\rm in}=0.0$ and $-0.5$; black circles, blue squares, and orange triangles denote fits with the solar, mild, and heavy grids, respectively. Filled symbols mark matched mixtures, open symbols mark mismatches, and the grey bar in panel~(c) gives the median 68\% posterior interval for the 54 fits.}
\label{fig:cno_mismatch}
\end{figure*}

Figure~\ref{fig:cno_mismatch} shows the complete cross-grid experiment. Filled symbols identify the matched solutions, while open symbols show fits with an incorrect CNO grid; the mean metallicity displacements for each mismatch are tabulated in Appendix~\ref{app:mock_diagnostics}. The direction of the solar--mild displacement was consistent at both input metallicities: fitting solar-generated spectra with the mild grid lowered the recovered \([Z]\) by approximately \(0.05\)--\(0.06\) dex, whereas fitting mild-generated spectra with the solar grid raised it by approximately \(0.05\) dex. This comparison supports a characteristic zero-point sensitivity of approximately \(0.04\)--\(0.05\) dex. For comparison, the median individual 68\% metallicity uncertainty for the observed Per~OB1 spectra is approximately \(0.18\)~dex with
the diagonal likelihood and \(0.20\)~dex with the regularised covariance likelihood (Sect.~\ref{sec:perob1_results}), substantially
larger than these shifts. A coherent zero-point displacement, however, does not average away across a stellar sample.

The CNO sensitivity increased towards higher temperature: the mean range in recovered metallicity across fitted grids rose from approximately 0.03~dex at 3600~K, through 0.05~dex at 3800~K, to 0.07--0.08~dex at 4000~K. The heavy mixture did not behave as a linear continuation of the mild case because the abundance changes interact non-linearly in spectral space. The mismatch was consequently redistributed among several correlated parameters rather than appearing only as a metallicity shift.

We interpret the approximately 0.04--0.05~dex solar--mild displacement
as a discrete model-choice sensitivity, not as a Gaussian random error to be added automatically in quadrature. Because it acts 
coherently, it does not decrease merely by combining several stars. We therefore adopted the physically motivated mild mixture for Per OB1, used the solar and heavy grids to bracket the association-level sensitivity, and did not interpret any grid as an individual CNO measurement.

\subsection{Relation to the Fisher diagnostics}
\label{sec:mocks_fisher_link}

The mock experiments show qualitative behaviour consistent with the parameter couplings identified by the Fisher diagnostics in Sect.~\ref{sec:fisher}. Lower recovered metallicities were generally accompanied by lower effective temperatures and higher microturbulent velocities, consistent with the signs of the local $T_{\rm eff}$--$[Z]$ and $\xi$--$[Z]$ covariance directions. In the full inference, the gravity prior substantially modifies the covariance geometry, as quantified below.

We also compared the local Fisher covariance quantitatively with the posterior covariance recovered from a separate set of six mocks at the primary Fisher nodes, which we refer to as the matched-configuration mocks. Their configuration was identical to that of the Fisher calculation in wavelength selection, continuum treatment, resolution, sampling, noise model, and gravity prior. The resolving power was fixed at $R = 2000$, so that the sampled parameter set was exactly the four-parameter set of the Fisher calculation. We then constructed a comparison Fisher matrix by adding the same $\sigma_{\log g} = 0.20$\,dex gravity prior used in the inference. The resulting prior-informed covariance can differ substantially from the spectral-only covariance shown in Fig.~\ref{fig:fisher_correlation}. We therefore retain the conventional spectral Fisher matrix in Sect.~\ref{sec:fisher} for the information-content diagnostics and use the comparison matrix only for the direct comparison with the posterior.

Across the six nodes, the ratios of the posterior 68\% half-width to the corresponding Fisher standard deviation range from 0.89 to 1.12 for $T_{\rm eff}$, 0.95 to 1.02 for $\log g$, 0.87 to 0.98 for $\xi$, and 0.81 to 1.00 for $[Z]$. The local Fisher approximation thus reproduces the characteristic uncertainty scale closely over the full set of representative nodes. The agreement is particularly close for $T_{\rm eff}$ and $\log g$, while the posterior uncertainties in $\xi$ and $[Z]$ are in some cases modestly narrower than the local prediction.

The principal covariance directions are also reproduced well. For the representative node ($T_{\rm eff} = 3800$\,K, $\log g = 0.0$, $\xi = 3$\,km\,s$^{-1}$, $[Z] = 0.0$), the comparison Fisher matrix gives $\rho(T_{\rm eff}, \log g) = 0.21$ and $\rho(\xi, [Z]) = -0.72$, against $0.19$ and $-0.70$ in the posterior. Across all six nodes, these dominant correlations follow the same node-to-node behaviour in the Fisher and Bayesian calculations. Some marginal widths and weaker correlations differ more noticeably, because the full posterior explores a finite and non-linear region of parameter space and is also subject to the boundaries of the model grid. At the node with the largest posterior-to-Fisher discrepancy, imposing the same grid boundaries on the Gaussian Fisher approximation accounts for a substantial part of the narrowing. These differences do not alter the main degeneracy structure identified by the local analysis.

This comparison therefore provides a direct check that, when the Fisher calculation and the Bayesian inference share the same wavelength selection, continuum treatment, parameter set, and gravity prior, the local Fisher matrix captures both the characteristic uncertainty scale and the principal covariance structure of the inference. The small nuisance-orthogonal fraction does not prevent successful metallicity recovery: although most of the local spectral response to metallicity resembles some combination of the responses to the other atmospheric parameters, the distinct component can still be strong enough to recover the correct metallicity when the full spectrum is fitted jointly.

Successful recovery with the correct grid does not, however, validate the absolute metallicity scale. Continuum placement or an incorrect abundance pattern can move the inferred solution coherently along the same coupled directions while retaining an acceptable fit. These systematic sensitivities are therefore not encoded by the posterior width alone and remain the subject of the controlled tests above and of Appendix~\ref{app:mock_diagnostics}.

\section{Application to Per OB1}
\label{sec:perob1}

\subsection{Stellar and association results}
\label{sec:perob1_results}

We applied the adopted analysis to 21 SpeX spectra representing 20 unique RSG targets towards Per OB1. We used the final wavelength selection, Gazak-like continuum normalisation, luminosity-based gravity prior, and both likelihood prescriptions described in Sect.~\ref{sec:inference}. Table~\ref{tab:perob1_results} summarises the association-level results, while Appendix~\ref{app:perob1} gives the individual posterior results and the detailed robustness checks. The association summaries include all 20 unique targets. We counted the two observations of BD+59~372 only once by averaging their posterior medians and uncertainties within each analysis configuration.

The mildly processed CNO grid with the regularised global covariance provided our preferred reference result. This grid approximates the carbon depletion and oxygen abundance measured in Per OB1 red supergiants, although it does not reproduce the composition of every star individually \citep{Fanelli2022}. The corresponding association mean was
\[
\langle[Z]\rangle=-0.095~{\rm dex},
\]
which we quote as approximately $-0.10$~dex. Its star-to-star dispersion was 0.09~dex, corresponding to a formal sampling error of 0.02~dex on the arithmetic mean, $s_{[Z]}/\sqrt{20}$. This error describes the sampling of the stellar point estimates and does not include systematic sensitivity shared by the sample.

The median individual metallicity uncertainty in the preferred configuration was approximately 0.20~dex, compared with 0.18~dex for the diagonal likelihood. These posterior intervals describe the precision of an individual spectrum and are distinct from both the observed dispersion and the systematic sensitivity of the common metallicity zero point.

\begin{table}
\caption{Association-level metallicity results for the Per~OB1 sample.}
\label{tab:perob1_results}
\centering
\begin{tabular}{llccc}
\hline\hline
CNO grid & Likelihood &
$\langle[Z]\rangle$ &
$s_{[Z]}$ &
$\widetilde{\sigma}_{[Z]}$ \\
 & & (dex) & (dex) & (dex) \\
\hline
Solar & Diagonal   & $-0.048$ & $0.092$ & $0.183$ \\
Solar & Covariance & $-0.067$ & $0.083$ & $0.190$ \\
Mild  & Diagonal   & $-0.091$ & $0.102$ & $0.180$ \\
Mild  & Covariance & $-0.095$ & $0.089$ & $0.195$ \\
Heavy & Diagonal   & $-0.131$ & $0.104$ & $0.180$ \\
Heavy & Covariance & $-0.105$ & $0.093$ & $0.203$ \\
\hline
\end{tabular}
\tablefoot{The columns give the unweighted mean, the sample standard deviation, and the median individual 68\% posterior uncertainty, with each asymmetric interval represented by the mean of its two half-widths. All statistics use 20 unique stars; the two results for BD+59~372 were averaged before calculating the summaries.}
\end{table}

\subsection{Robustness of the metallicity scale}
\label{sec:perob1_robustness}

The six combinations of CNO mixture and likelihood produced association means between $-0.05$ and $-0.13$~dex. At fixed CNO mixture, changing the likelihood altered the mean by at most 0.03~dex and by less than 0.01~dex for the preferred mild grid. Relative to the solar-CNO result, the mild and heavy grids shifted the covariance-based means by 0.03 and 0.04~dex, respectively, compared with 0.04 and 0.08~dex for the diagonal likelihood. The regularised covariance therefore reduced the dependence on the assumed CNO mixture, particularly for the deliberately extreme heavy grid, while supporting the characteristic 0.04--0.05~dex abundance-pattern sensitivity found in the mocks.

In the diagonal heavy-grid analysis, HD 13136 reached the lower
model-grid boundary at $T_{\rm eff}=3400$~K, whereas the covariance
redistributed the mismatch among $T_{\rm eff}$, $\xi$, and $[Z]$ and
moved the solution away from the boundary. BD+56 724 showed the strongest
structured residuals, but excluding it from the preferred configuration
changed the mean from $-0.095$ to $-0.100$~dex and left the dispersion
unchanged at $0.09$~dex. Excluding the three targets with the least secure
association membership, BD+56 724, BD+58 501, and HD 236947, changed the
preferred mean from $-0.095$ to $-0.096$~dex and likewise left the
dispersion unchanged at $0.09$~dex. None of these checks therefore changes
the adopted association result.

We repeated the preferred analysis with the broad fixed priors $\log g\sim\mathcal{N}(0.0,0.5^2)$ and $\log g\sim\mathcal{N}(0.2,0.5^2)$. Relative to the luminosity-based prior, these choices lowered the association mean by 0.03~dex and by less than 0.01~dex, respectively, although the typical posterior gravity uncertainty approximately doubled. 
The paired shifts followed the coupled Fisher directions, with
Pearson correlation coefficients of \(r=0.98\) between
\(\Delta T_{\rm eff}\) and \(\Delta\log g\), and \(r=0.97\) between
\(\Delta[Z]\) and \(\Delta\log g\). The luminosity-based prior therefore regularises the weakest, gravity-dominated mode without determining the association metallicity scale.

\subsection{Repeatability and external checks}
\label{sec:perob1_checks}
The two observations of BD+59~372 differed by \(-0.04\)~dex in
metallicity in the preferred analysis. Three programme targets,
HD~14404, HD~14469, and HD~14488, also have independent spectra in
the IRTF Spectral Library \citep{Rayner2009}. Analysed with the same
mild-CNO covariance configuration, these archival spectra gave
archival-minus-programme metallicity differences of \(-0.08\),
\(-0.01\), and \(-0.02\)~dex, respectively. Their mean difference was
\(-0.04\)~dex, their RMS difference was \(0.05\)~dex, and the maximum
absolute difference was \(0.08\)~dex. All of these differences are smaller than the corresponding
individual 68\% metallicity uncertainties, which are typically
about 0.2 dex; Appendix~\ref{app:perob1} gives the parameter-level comparisons.

As a further external check, we directly fitted the native-resolution
$J$-band spectra of V424~Lac and Betelgeuse from the Measuring
Abundances of red super Giants with Infrared Spectroscopy (MAGIS) project 
\citep[R=28000,][]{Taniguchi2025} and compared the results with fits to their
corresponding SpeX spectra from the IRTF Spectral Library. The atmospheric 
parameters inferred from the two
datasets agree closely for V424~Lac. For Betelgeuse, surface gravity
and microturbulence likewise agree well, while the effective
temperatures differ by approximately \(160\)~K. The resulting
SpeX-minus-MAGIS metallicity differences are \(-0.08\) and
\(+0.03\)~dex for V424~Lac and Betelgeuse, respectively.

The SpeX results are also broadly consistent with the independent
analysis of the MAGIS spectra by \citet{Taniguchi2025}. For V424~Lac
and Betelgeuse, respectively, the SpeX temperatures differ from their
values by \(+25\) and \(-34\)~K, and the surface gravities by
\(+0.06\) and \(+0.02\)~dex. The corresponding differences between
our global \([Z]\) coordinate and their \([\mathrm{Fe/H}]\) values are
approximately \(0.00\) and \(+0.02\)~dex, although these abundance
quantities are not strictly equivalent. Our SpeX microturbulent
velocities are higher by approximately \(0.8\) and
\(1.6\,{\rm km\,s^{-1}}\), so the agreement does not extend uniformly
to every fitted parameter.

Table~\ref{tab:repeatability_comparisons} gives the corresponding
parameter-level differences for the programme repeat, the three IRTF
Spectral Library pairs, and the two MAGIS comparisons. Taken together, 
the six comparisons support the internal
repeatability of the metallicity inference at a level below the
typical individual uncertainty. They do not, however,
establish an external abundance calibration.

 \section{Discussion}
\label{sec:discussion}

\subsection{Information, precision, and robustness}
\label{sec:discussion_information}

The extended J-band spectra respond measurably to all four fitted atmospheric parameters, but much of this sensitivity lies along coupled directions. Only a small fraction of the raw metallicity information remains independent after projecting out the effective-temperature, surface-gravity, and microturbulence directions, while surface gravity dominates the weakest information mode. A physically motivated gravity prior is therefore an appropriate regularising constraint for the inference. These results distinguish the detection of a spectral response from the independent identification of its physical origin.

This distinction has direct consequences for observational design and the interpretation of posterior uncertainties. At fixed wavelength coverage and relative pixel weighting, a uniform increase in $\mathrm{S/N}$ rescales the Fisher matrix but does not change the normalised response geometry or the independent-information fractions. It can tighten the posterior along parameter combinations to which the spectrum is already sensitive, but cannot separate response directions with similar shapes. Additional wavelength coverage can improve spectroscopic identifiability, whereas an external gravity prior improves the conditioning of the posterior without creating new spectroscopic information. The external gravity constraint helps the fit where the spectrum alone cannot pin down the surface gravity well. As a result, the reported surface gravities are prior-informed
rather than purely spectroscopic measurements.

The gravity-prior experiment provides an empirical test of this interpretation. Replacing the adopted luminosity-based prior with either of the two deliberately broad Gaussian priors changed the Per OB1 mean metallicity by at most 0.03~dex, although the typical posterior uncertainty in $\log g$ approximately doubled. The paired parameter shifts followed the coupled $T_{\rm eff}$--$\log g$ and $[Z]$--$\log g$ directions identified by the Fisher analysis. The external prior therefore constrains the location of the solution along a weakly identifiable direction without determining the association metallicity scale.

The controlled experiments establish a clear hierarchy among the
analysis sensitivities tested here. For the representative mock setup,
standard continuum treatment underestimated the metallicity by approximately 0.23~dex, whereas the model-guided Gazak-like treatment reduced the mean bias to approximately $-0.01$~dex. Plausible CNO processing introduced a smaller metallicity-scale sensitivity of approximately 0.04--0.05~dex, and changing from the diagonal to the regularised-covariance likelihood altered the Per OB1 association mean by at most 0.03~dex for a fixed CNO grid. Continuum placement is therefore the largest demonstrated vulnerability in the present analysis, while the CNO mixture and treatment of correlated residuals contribute more modestly to the zero point.

The same distinction applies when combining several stars. Averaging reduces the statistical uncertainty of the association mean and suppresses star-specific fluctuations, but it does not remove a continuum or abundance-pattern sensitivity shared by the full sample. Likewise, a narrow posterior quantifies precision conditional on the adopted forward model, likelihood, normalisation, and priors, but does not by itself demonstrate the robustness of the absolute metallicity scale. Reliable applications therefore require continuum and mismatch tests matched to the wavelength coverage, resolution, and $\mathrm{S/N}$ of the observations, rather than relying on posterior precision alone.

\subsection{The metallicity of Per OB1}
\label{sec:discussion_perob1}

The physically motivated mild-CNO grid combined with the regularised global covariance gives a mean metallicity of approximately $-0.10$~dex for Per OB1. Within the regularised-covariance analysis, the solar, mild, and heavy-CNO results span approximately 0.04~dex; across all six combinations of CNO grid and likelihood, the complete range is approximately 0.08~dex and is driven mainly by the diagonal heavy-grid result. The abundance-pattern and residual-covariance choices tested here therefore do not by themselves produce a shift comparable to the separation from the lower abundance scale reported by \citet{Fanelli2022}. Within the adopted LTE MARCS/TURBOSPECTRUM framework, our result indicates that the global metallicity coordinate of Per OB1 is mildly rather than strongly subsolar.

A comparison of association means alone cannot identify the origin
of the different literature scales. Based on high-resolution
\(J\)-band spectra and an LTE grid, \citet{Gazak2014} obtained
\([Z]=+0.04\pm0.10\)~dex for 11 RSGs in Per~OB1, where the quoted
uncertainty is the star-to-star standard deviation. The LTE analysis
of \citet{Fanelli2022}, based on high-resolution optical and
near-infrared spectroscopy of 27 RSGs in the wider Perseus complex,
instead found a mean iron abundance of approximately
\([\mathrm{Fe/H}]=-0.30\)~dex. Fanelli et al. adopted
\(\epsilon_{\odot}(\mathrm{Fe})=7.50\), compared with 7.45 in our
model mixture, so their mean becomes approximately
\([\mathrm{Fe/H}]=-0.25\)~dex when expressed relative to our solar
iron reference. This adjustment reduces, but does not remove, the
separation between the abundance scales.

The common-star comparisons show that the atmospheric-parameter scales also differ systematically. For the eight stars shared with \citet{Fanelli2022}, our preferred solutions are hotter by an average of 134~K and have surface gravities higher by 0.24~dex, while the mean microturbulence difference is only $+0.06$~km~s$^{-1}$. After harmonising the solar iron reference, the mean paired difference between our global metallicity coordinate and their near-infrared Fe\,{\sc i} abundances is
\[
\left\langle
[Z]_{\rm SpeX}
-
[\mathrm{Fe/H}]_{\rm Fanelli}
\right\rangle
=
+0.21~{\rm dex},
\]
with a star-to-star scatter of 0.10~dex. The gravity comparison should be interpreted cautiously because our gravities were regularised by the luminosity-based prior, whereas Fanelli et al. calculated their gravities from luminosities and an adopted stellar mass.

For the 11 stars shared with \citet{Gazak2014}, comparison with
their LTE solutions gives mean SpeX-minus-Gazak differences of
approximately \(-190\)~K in \(T_{\rm eff}\), \(0.00\)~dex in
\(\log g\), \(-0.20\)~km~s\(^{-1}\) in \(\xi\), and
\(-0.12\)~dex in metallicity. The metallicity differences have a
star-to-star scatter of 0.12~dex, indicating both a systematic
offset and appreciable variation among individual objects. For
the three stars common to all samples, our temperatures are on
average approximately 170~K hotter than those of Fanelli et al.
and 150~K cooler than the LTE results of Gazak et al. Our
temperature scale therefore lies between the cooler Fanelli and
hotter Gazak LTE scales for the directly comparable stars.

Diagnostic-dependent temperature scales are well established for
RSGs. \citet{Davies2013} found that temperatures obtained from
optical TiO bands were systematically cooler than those inferred
from line-free optical--near-infrared spectral energy distributions
and flux integration. They associated this difference with the
sensitivity of molecular features to the temperature structure of
the outer atmosphere, and excluded the CO bands from their
continuum-based temperature determination for the same reason.
The temperatures of \citet{Fanelli2022} are instead based on the
balance between carbon abundances inferred from atomic
C~{\sc i} and molecular CO lines, following the C-thermometer
introduced by \citet{Fanelli2021}. Their method is therefore
independent of the TiO diagnostic, but its molecular component may
remain sensitive to the atmospheric temperature structure. The
difference found here could consequently include a
diagnostic-dependent temperature zero point, although the available
comparisons do not establish its origin or magnitude.

Within our extended \(J\)-band analysis, the local information
geometry provides a quantitative context for the possible effect of
this temperature difference on metallicity.  For the eight stars shared with \citet{Fanelli2022}, applying the
prior-informed local Fisher compensation of $+0.059$ to $+0.095$~dex in
$[Z]$ per $+100$~K to the mean temperature offset of $+134$~K gives a
metallicity displacement of approximately $+0.08$ to $+0.13$~dex. This is
in the same direction as the measured $+0.21$~dex difference, showing that
the temperature-scale offset could plausibly contribute a substantial
fraction, but not all, of the separation. This estimate is diagnostic
rather than a correction because the Fisher response is local and varies
with atmospheric parameters. Temperature alone therefore cannot provide a unique explanation,
particularly because the analyses also differ in wavelength coverage,
spectral resolution, continuum treatment, abundance pattern, atomic
and molecular data, gravity determination, and the specific spectral
diagnostics used.

Independent results for hot supergiants provide a useful local comparison. Eight B- and A-type supergiants associated with Per OB1 and analysed with non-LTE methods by \citet{Firnstein2010}, \citet{Wessmayer2022}, and \citet{Wessmayer2023} give $\langle[\mathrm{Fe/H}]\rangle=-0.06$~dex relative to the solar iron abundance of 7.45 adopted in our model mixture, with a star-to-star dispersion of approximately 0.07~dex. An inverse-variance weighted calculation gives the same abundance to the quoted precision, $-0.06\pm0.03$~dex. For the three stars with published metal mass fractions, the mean $Z=0.0113$ corresponds to $[Z]\simeq-0.09$~dex for $Z_{\odot}=0.014$. Although the BA-supergiant iron abundance and our RSG grid coordinate are not identical quantities, both the absolute iron abundance and the independently tabulated metal mass fractions support a mildly subsolar present-day composition in Per OB1.

\subsection{Scope and limitations}
\label{sec:discussion_limitations}

Our conclusions apply specifically to the extended $1.155$--$1.340\,\mu{\rm m}$ interval analysed at $R\simeq2000$. In the spectral information-density curves calculated before adding the gravity prior, the classical $1.155$--$1.220\,\mu{\rm m}$ interval contains median fractions of 40\% of the raw and 34\% of the nuisance-orthogonal metallicity information across the six principal nodes. The red extension therefore supplies median fractions of 60\% and 66\%, respectively, with important contributions near the Al\,{\sc i} features at approximately $1.312$--$1.315\,\mu{\rm m}$. The present information content and recovery performance should not be transferred without qualification to the classical restricted interval.

The absolute abundance scale remains conditional on the
one-dimensional MARCS atmospheres, the adopted atomic and molecular
data, and the LTE line-formation treatment used throughout the solar,
mild-CNO, and heavy-CNO grids. We did not apply non-LTE corrections to
any of the production analyses; non-LTE 
calculations predict species- and parameter-dependent effects for
several important J-band transitions
\citep{Bergemann2012,Bergemann2013,Bergemann2015}, and the LTE and
non-LTE association means reported by \citet{Gazak2014} for Per OB1
differed by approximately 0.08~dex. This displacement cannot be
applied directly as a correction because the wavelength coverage, line
selection, and modelling framework differ. The RSG analyses of \citet{Fanelli2022} and \citet{Taniguchi2025} were likewise performed in LTE, so comparisons with those studies do not test the effect of LTE versus non-LTE line formation. Moreover, the three fixed CNO mixtures used here modify the emergent spectra without recalculating the atmospheric structures and do not sample a continuous distribution of evolutionary abundance patterns.

Finally, Per~OB1 represents one association with finite membership
uncertainties. The repeatability sample comprises one programme repeat
and three IRTF-library pairs, while the external MAGIS comparison
contains only two stars. The present work therefore assesses information content and robustness within a controlled local application rather than establishing a universal RSG abundance calibration. Extragalactic applications should retain the extended red coverage where possible, validate continuum placement with realistic mocks, constrain weak atmospheric directions externally, and test abundance-pattern and residual-covariance sensitivity using the observing configuration of the science data. Applications to unresolved RSG-dominated clusters require additional validation because stellar mixtures and internal velocity broadening introduce effects not examined here.

\section{Conclusions}
\label{sec:conclusions}

We investigated the information content and robustness of extended J-band spectroscopy of RSGs at $R\simeq2000$. Our main conclusions are as follows.

\begin{enumerate}
    \item The spectra respond measurably to $T_{\rm eff}$, $\log g$, $\xi$, and $[Z]$, but these responses are strongly coupled. Only approximately 8--10\% of the raw metallicity information and 4--7\% of the raw surface-gravity information remain independent of the other fitted directions, while surface gravity dominates the weakest Fisher mode. An external gravity prior improves the conditioning of the inference
by constraining the weak gravity-dominated direction, so the resulting surface gravities are prior-informed rather than purely spectroscopic
measurements.

    \item The wavelength extension beyond the classical $1.155$--$1.220\,\mu{\rm m}$ interval
makes a substantial contribution to the metallicity inference, roughly in proportion to its larger wavelength coverage. Across the six principal Fisher nodes, it supplies median fractions of 60\% of the raw and 66\% of the nuisance-orthogonal metallicity information available over the complete $1.155$--$1.340\,\mu{\rm m}$ range.

    \item In the matched-grid mock observations, the full Bayesian inference recovered the injected metallicities with an RMS error of approximately 0.04~dex and a maximum absolute offset of 0.06~dex. Coupled local information geometry therefore does not preclude reliable recovery when the forward model and analysis assumptions are internally consistent.

    \item Continuum normalisation is the largest systematic
    vulnerability demonstrated by our tests. For representative
    solar-metallicity mocks, the standard continuum treatment produced a metallicity bias of approximately $-0.23$~dex, whereas the adopted model-guided Gazak-like method reduced the mean bias to approximately $-0.01$~dex.

    \item Plausible CNO processing introduces a discrete model-choice sensitivity of approximately 0.04--0.05~dex in the metallicity scale. For Per OB1, the regularised global covariance reduced the solar-to-heavy grid displacement in the association mean from approximately 0.08 to 0.04~dex, indicating that part of the extreme-grid mismatch can remain as structured residuals rather than being projected entirely onto the atmospheric parameters.

    \item For the 20 unique RSG targets towards Per OB1, the physically motivated mild-CNO grid combined with the regularised global covariance gives a mean metallicity of approximately $-0.10$~dex, with an observed star-to-star dispersion of approximately 0.09~dex. This result supports a mildly subsolar composition for the association within the adopted one-dimensional LTE MARCS/TURBOSPECTRUM framework, while the absolute scale remains conditional on the continuum treatment, fixed abundance pattern, line data, and model atmospheres.

\end{enumerate}

Extended $J$-band spectroscopy therefore offers more than a
metallicity estimate: the present framework also identifies which part of that estimate is supplied by the spectrum and which remains conditional on external information and modelling choices. The wavelength range and resolution 
considered here are particularly well matched to future applications with JWST/NIRSpec and ELT/MOSAIC. NIRSpec provides access to the extended J band at resolving powers of order $10^3$--$3\times10^3$, enabling quantitative spectroscopy of individual RSGs in nearby galaxies, while ELT-class facilities should extend such work to distances of order 10~Mpc and potentially to tens of Mpc, subject to crowding \citep{Evans2011}.

Extending the analysis beyond $1.34\,\mu{\rm m}$ is also a natural next step, since both instruments provide substantially broader near-infrared coverage. Whether 
those wavelengths add genuinely independent atmospheric information, rather than predominantly redundant sensitivity or additional modelling systematics, requires
the same information-content and recovery tests applied here. Nearby systems such as Per~OB1 therefore provide the bridge between
methodological validation and applications to increasingly distant stellar populations. That bridge, rather than formal precision alone,
is what gives the technique its value as an extragalactic abundance diagnostic.

\section*{Data availability}

The MARCS model atmospheres, the TURBOSPECTRUM spectral-synthesis code,
the Vienna Atomic Line Database (VALD) atomic data, and the molecular line lists used in this work are
publicly available through their respective repositories. The IRTF Spectral
Library spectra and the MAGIS project spectra used for the external
comparisons are likewise publicly available from the sources cited in
Sect.~5.3. The reduced SpeX spectra covering the extended \(J\)-band interval
for the 21 observations of 20 RSG targets towards Per~OB1 are
available from the corresponding author upon reasonable request.

\begin{acknowledgements}

This work is based on observations obtained at the NASA Infrared Telescope Facility, which is operated by the University of Hawai{\textokina}i under contract 80HQTR24DA010 with the National Aeronautics and Space Administration.

The authors wish to recognise and acknowledge the very significant cultural role and reverence that the summit of Maunakea has always had within the Native Hawaiian community. We are most fortunate to have the opportunity to conduct observations from this mountain.

RPK acknowledges support by the Munich Excellence Cluster Origins and the Munich Institute for Astro-, Particle and Biophysics (MIAPbP) both funded by the Deutsche Forschungsgemeinschaft (DFG, German Research Foundation) under the German Excellence Strategy EXC-2094 390783311.

We thank the referee for a careful and constructive report that helped
to improve the clarity and presentation of the manuscript.

\end{acknowledgements}

\bibliographystyle{aa} 
\bibliography{biblio}

\begin{appendix}
\nolinenumbers   

\section{Synthetic spectral framework and scope}
\label{app:grid}

This appendix summarises the properties of the synthetic grids that are most relevant to the interpretation of Sect.~\ref{sec:models}. The spectral preparation and MAUI emulator construction and validation followed \citet{Urbaneja2026} and are not repeated here. The atomic line list was taken from VALD
\citep{Kupka1999,Ryabchikova2015}, while the molecular data were adopted
from the compilation by B. Plez\footnote{https://www.lupm.in2p3.fr/users/plez/}. No empirical recalibration of
oscillator strengths was performed; all transition probabilities were used as provided by the original sources. 

\subsection{MARCS atmosphere grid}
\label{app:marcs_grid}

The atmospheric structures were drawn from the dedicated MARCS grid for red supergiants \citep{Gustafsson2008}. These one-dimensional, hydrostatic, spherical models adopted a stellar mass of $15\,M_\odot$, the 
scaled solar composition of \citet{Grevesse2007}, and an atmospheric microturbulence of $\xi_{\rm atm}=5\,{\rm km\,s^{-1}}$. We used the calculated structures at the nodes specified in Sect.~\ref{sec:models} and did not interpolate between atmospheric structures when constructing the synthetic spectral grids. The emulator subsequently provided continuous spectral predictions within the resulting parameter domain; it did not interpolate or modify the MARCS structures themselves.

The fixed mass sets the radius and geometrical extension of a spherical atmosphere at a given gravity. We did not investigate sensitivity to this choice, so the grid represents a spectral interpolation domain constructed from a specified family of MARCS atmospheres rather than an evolutionary sequence in which every parameter combination lies on a $15\,M_\odot$ track.

\subsection{Microturbulence in the atmosphere and spectral synthesis}
\label{app:microturbulence}

Although all MARCS structures adopted $\xi_{\rm atm}=5\,{\rm km\,s^{-1}}$, the TURBOSPECTRUM calculations sampled the synthesis microturbulence from 0.5 to 6.5~${\rm km\,s^{-1}}$ in steps of 0.5~${\rm km\,s^{-1}}$. The synthesis therefore includes the direct influence of microturbulence on line opacity and saturation, but not the corresponding response of the atmospheric structure, including turbulent pressure and geometrical extension. The inferred $\xi$ should consequently be interpreted as the line-formation microturbulence required within the fixed MARCS structure; models with $\xi\neq\xi_{\rm atm}$ are not fully self-consistent in their treatment of microturbulence.

\subsection{CNO abundance mixtures}
\label{app:cno_grids}

The MARCS structures retained the scaled-solar abundance pattern at every atmospheric node, while the three CNO mixtures listed in Sect.~\ref{sec:models} were introduced only during the TURBOSPECTRUM calculations. TURBOSPECTRUM recomputed the chemical and molecular equilibrium for each mixture, so the spectra include the direct effects of the modified abundances on atomic and molecular opacity and on the local pseudo-continuum. We did not recompute the atmospheric temperature and pressure structures, however, and the grid comparisons therefore exclude structural feedback from the altered CNO composition.

The three mixtures are discrete modelling alternatives rather than samples from a continuous distribution of RSG surface compositions. Their differences quantify model-choice sensitivity within the tested patterns; they do not provide individual CNO measurements or define an automatically Gaussian contribution to the metallicity uncertainty. We likewise trained and evaluated the three emulators separately, without interpolation between CNO mixtures.

\subsection{Scope of the model framework}
\label{app:model_scope}

The grids were designed to quantify parameter sensitivity and recovery within one explicitly defined family of synthetic spectra. They include the direct spectral responses to $T_{\rm eff}$, $\log g$, synthesis microturbulence, global metallicity, and three fixed CNO patterns. We did not explore alternative atmospheric geometries, stellar masses, convection prescriptions, solar abundance scales, line lists, or self-consistent structures for different microturbulent velocities and CNO mixtures.

All calculations assumed one-dimensional hydrostatic atmospheres and
LTE line formation. They therefore omit non-LTE effects, time-dependent velocity fields, large-scale convective inhomogeneities, and other three-dimensional phenomena expected in RSG atmospheres. Although some detailed signatures are blended at $R\simeq2000$, their effects on line strengths and the pseudo-continuum need not vanish and remain part of the model dependence of the absolute metallicity scale.

Emulator validation tests whether the numerical approximation reproduces the calculated TURBOSPECTRUM spectra, whereas the matched-grid mocks test parameter recovery when the generating spectra belong to the same model family. The mismatched-grid experiments additionally test the selected continuum and CNO alternatives, but none of these calculations samples every possible limitation of the atmosphere models, line data, or LTE approximation. The Per~OB1 result should consequently be interpreted as the metallicity supported by the extended $J$-band spectra within the specified MARCS/TURBOSPECTRUM framework rather than as a model-independent absolute abundance scale.

\section{Fisher-diagnostic construction and validation}
\label{app:fisher}

This appendix gives the numerical setup and definitions required to interpret the Fisher diagnostics in Sect.~\ref{sec:fisher}, following the general formalism of \citet{Tegmark1997}, \citet{Coe2009}, and \citet{Heavens2014}. It also documents the finite-difference stability tests and node-by-node results underlying the figures and values quoted in the main text.

\subsection{Diagnostic construction}
\label{app:fisher:setup}
\label{app:fisher:definitions}

We calculated the diagnostics over the three intervals $1.155$--$1.23\,\mu{\rm m}$, $1.23$--$1.285\,\mu{\rm m}$, and $1.285$--$1.34\,\mu{\rm m}$. The spectra were degraded to $R=2000$, sampled with three pixels per resolution element, and assigned a constant $\mathrm{S/N}=200$ per pixel. 

We adopted the diagonal covariance $C={\rm diag}(\sigma_i^2)$ with $\sigma_i=1/200$ in continuum-normalised flux units. We did not include the regularised Mat\'ern-$3/2$ covariance used for the observed spectra because the Fisher calculation was designed to isolate the local spectral-response geometry under an idealised noise model. Its formal uncertainties are therefore not predictions of the final posterior precision.

Before constructing each derivative, we applied the same model-guided continuum profiling used in the spectral inference 
(see Appendix~\ref{app:continuum}). Within each window, we selected the 15\% of valid pixels whose displaced-model fluxes lay closest to unity and fitted the logarithmic ratio of the displaced model to the fiducial spectrum at those anchors with a constant-plus-linear function. We then applied the corresponding multiplicative correction to the displaced model. The anchors and coefficients were recomputed for every displaced spectrum, no prior was imposed on the continuum coefficients, and no additional projection over polynomial modes was applied. Thus, ``raw information'' below denotes the single-parameter information after masking and continuum profiling but before projection over the other atmospheric derivatives.

For $\boldsymbol{\theta}=(T_{\rm eff},\log g,\xi,[Z])$, we defined
\[
J_{i\alpha}
=
\frac{\partial f_i}{\partial\theta_\alpha},
\qquad
F_{\alpha\beta}^{\rm spec}
=
J_\alpha^{\rm T}C^{-1}J_\beta.
\]
With the noise-weighted derivative $\boldsymbol{d}_\alpha=C^{-1/2}\boldsymbol{J}_\alpha$, the diagonal element is $F_{\alpha\alpha}^{\rm spec}=\boldsymbol{d}_\alpha^{\rm T}\boldsymbol{d}_\alpha$, while the off-diagonal elements measure overlaps among the spectral responses.

For eigenmode and condition-number calculations, we used the scaled coordinates
\[
\boldsymbol{q}
=
D^{-1}(\boldsymbol{\theta}-\boldsymbol{\theta}_0),
\qquad
D
=
{\rm diag}
\left(
100\,{\rm K},
0.1\,{\rm dex},
1\,{\rm km\,s^{-1}},
0.1\,{\rm dex}
\right),
\]
for which
\[
F^{(q)}
=
D^{\rm T}F^{\rm spec}D.
\]
We diagonalised
\[
F^{(q)}V
=
V\Lambda,
\]
ordered the modes from the largest to the smallest eigenvalue, and defined $\kappa=\lambda_{\rm max}/\lambda_{\rm min}$. The overall sign of each eigenvector is arbitrary, so only the relative signs and amplitudes of its components are meaningful.

The spectral covariance approximation and its correlation coefficients were
\[
\Sigma_\theta^{\rm spec}
\simeq
\left(F^{\rm spec}\right)^{-1},
\qquad
\rho_{\alpha\beta}
=
\frac{
(\Sigma_\theta^{\rm spec})_{\alpha\beta}
}{
\sqrt{
(\Sigma_\theta^{\rm spec})_{\alpha\alpha}
(\Sigma_\theta^{\rm spec})_{\beta\beta}
}
}.
\]
The inverse Fisher matrix is the multiparameter Cramér--Rao lower bound on the covariance of unbiased estimators
\citep{Rao1945,Cramer1946}. The correlation panel in Fig.~\ref{fig:fisher_correlation} shows these marginalised covariance correlations rather than the pairwise cosines between derivatives, thereby including the simultaneous propagation of all four parameter couplings.

We included the gravity prior through
\[
F^{\rm post}
=
F^{\rm spec}
+
F^{\rm prior}.
\]
Its only non-zero element in physical coordinates was $F_{\log g,\log g}^{\rm prior}=1/\sigma_{\log g}^2$, with $\sigma_{\log g}=0.20$~dex, corresponding to
\[
F_{\log g,\log g}^{\rm prior,(q)}
=
\left(
\frac{0.1}{0.20}
\right)^2
=
0.25
\]
in the scaled coordinates. We calculated the eigenmodes, independent-information fractions, and wavelength-resolved quantities from the spectral matrix before adding the prior. We used the prior-informed matrix only for the covariance, condition-number, and temperature--metallicity compensation diagnostics reported in Sect.~\ref{sec:fisher}.

\subsection{Independent and wavelength-resolved information}
\label{app:fisher:eta}
For parameter \(\alpha\), we collected the noise-weighted derivatives
of the other three parameters into the nuisance-derivative matrix

\[
N_{-\alpha}
=
\begin{bmatrix}
d_{\beta_1} & d_{\beta_2} & d_{\beta_3}
\end{bmatrix},
\qquad \beta_k\neq\alpha .
\]

We denote by \(P_{-\alpha}\) the orthogonal projector onto the column
space of \(N_{-\alpha}\). Thus, \(P_{-\alpha}d_\alpha\) is the part
of the response to parameter \(\alpha\) that can be reproduced by a
linear combination of the other three parameter responses. The
remaining nuisance-orthogonal component is
\[
d_{\alpha,\perp}
=
\left(I-P_{-\alpha}\right)d_\alpha ,
\]
where \(I\) is the identity matrix. We then defined the
independent-information fraction as
\[
\eta_\alpha
=
\frac{d_{\alpha,\perp}^{\rm T}d_{\alpha,\perp}}
     {d_\alpha^{\rm T}d_\alpha}
=
\frac{1}
{F_{\alpha\alpha}^{\rm spec}
 \left[\left(F^{\rm spec}\right)^{-1}\right]_{\alpha\alpha}} .
\]
This least-squares projection is equivalent to marginalising over the
other parameters through the Schur complement of their block in the
Fisher matrix, a standard result in Fisher-information analysis
\citep[e.g.][]{Tegmark1997,Coe2009}. We evaluated the projection by
solving the least-squares problem directly rather than by explicitly
forming or inverting the nuisance-derivative overlap matrix; the final
equality holds when the spectral Fisher matrix has full rank. The
quantity \(\eta_\alpha\) is the fraction of the single-parameter
sensitivity that cannot be reproduced by a linear combination of the
other atmospheric responses, while \(1/\eta_\alpha\) is the
corresponding local variance-inflation factor.

The raw and nuisance-orthogonal information densities were
\[
I_{\alpha,i}^{\rm raw}
=
d_{i\alpha}^{2},
\qquad
I_{\alpha,i}^{\perp}
=
d_{i\alpha,\perp}^{2},
\]
with cumulative fractions
\[
{\cal C}_{\alpha}^{X}(\lambda)
=
\frac{
\sum_{\lambda_i\leq\lambda}I_{\alpha,i}^{X}
}{
\sum_i I_{\alpha,i}^{X}
},
\qquad
X\in\{{\rm raw},\perp\}.
\]
We calculated the nuisance projection globally over the complete wavelength interval before assigning its squared residual response to individual pixels. The resulting curves locate the globally distinct part of a parameter response and should not be interpreted as independent line-by-line abundance measurements.

\begin{figure}
\centering
\includegraphics[]{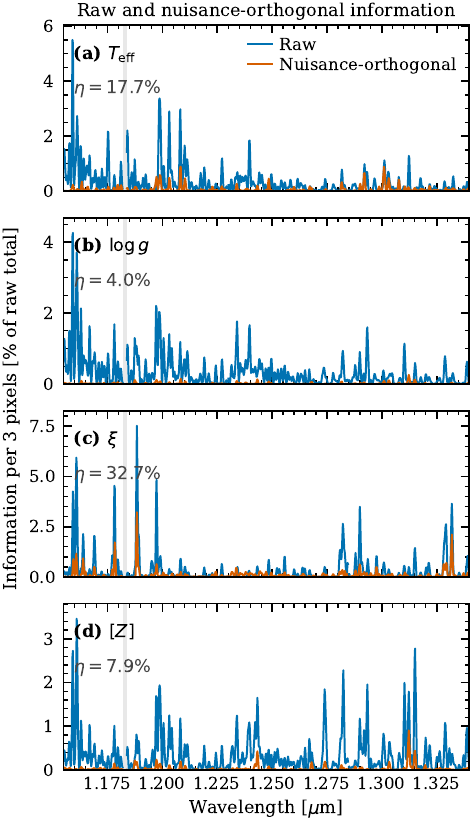}
\caption{Raw and nuisance-orthogonal information densities for the
representative $T_{\rm eff}=3800\,{\rm K}$, $\log g=0.0$, $\xi=3\,{\rm km\,s^{-1}}$, and $[Z]=0.0$ node. Each panel gives the information per three pixels as a percentage of the corresponding total raw information; the annotation reports the independent-information fraction $\eta_\alpha$.} 
\label{fig:fisher_information_density}
\end{figure}

Figure~\ref{fig:fisher_information_density} illustrates the raw and nuisance-orthogonal densities for all four parameters at the representative 3800~K solar-metallicity node. For metallicity, a prominent raw response occurs near $1.159$--$1.161\,\mu{\rm m}$, whereas important distinct contributions occur near $1.243\,\mu{\rm m}$ and $1.312$--$1.315\,\mu{\rm m}$. The distributions therefore do not simply follow absorption depth, and individual peaks generally reflect several transitions blended at $R=2000$.

\subsection{Finite differences and stability tests}
\label{app:fisher:stability}

We calculated each derivative from symmetric emulator evaluations according to
\[
\left.
\frac{\partial f}{\partial\theta_\alpha}
\right|_{\boldsymbol{\theta}_0}
\simeq
\frac{
f(\boldsymbol{\theta}_0+h_\alpha\boldsymbol{e}_\alpha)
-
f(\boldsymbol{\theta}_0-h_\alpha\boldsymbol{e}_\alpha)
}{
2h_\alpha
}.
\]
The primary and stability-test half-steps are listed in Table~\ref{tab:fisher_scaling}. Both sets of displaced spectra underwent identical degradation, resampling, masking, and continuum profiling.

\begin{table}
\caption{Parameter scaling units and finite-difference half-steps.}
\label{tab:fisher_scaling}
\centering
\begin{tabular}{lccc}
\hline\hline
Parameter &
Scaling unit &
Primary step &
Stability step \\
\hline
$T_{\rm eff}$ &
100~K &
50~K &
100~K \\
$\log g$ &
0.1~dex &
0.1~dex &
0.2~dex \\
$\xi$ &
1~km~s$^{-1}$ &
0.2~km~s$^{-1}$ &
0.4~km~s$^{-1}$ \\
$[Z]$ &
0.1~dex &
0.05~dex &
0.10~dex \\
\hline
\end{tabular}
\tablefoot{The steps are half-widths of the symmetric finite differences. The primary derivatives used the smaller values, while the larger values were used only for the stability tests.}
\end{table}

For each parameter and node, we compared the primary derivative with that obtained using the larger step through
\[
\Delta_{\rm RMS}
=
\frac{
{\rm RMS}
(\boldsymbol{d}_{\rm large}-\boldsymbol{d}_{\rm primary})
}{
{\rm RMS}(\boldsymbol{d}_{\rm primary})
},
\qquad
c
=
\frac{
\boldsymbol{d}_{\rm large}^{\rm T}
\boldsymbol{d}_{\rm primary}
}{
\lVert\boldsymbol{d}_{\rm large}\rVert
\lVert\boldsymbol{d}_{\rm primary}\rVert
}.
\]
The strict criteria were $\Delta_{\rm RMS}\leq0.05$ and $c\geq0.995$, while cases with $\Delta_{\rm RMS}\leq0.15$ and $c\geq0.98$ were accepted but flagged for inspection. These measures test the derivative amplitude and orientation, respectively.

Twenty-three of the 24 comparisons passed the strict criteria. The sole exception was the $T_{\rm eff}$ derivative at $T_{\rm eff}=3600$~K and $[Z]=-0.5$, for which $\Delta_{\rm RMS}=0.0506$ and $c=0.99884$; it passed the secondary criteria and preserved the weakest eigenmode and overall Fisher geometry. None of the adopted derivatives therefore required rejection.

\subsection{Node-by-node diagnostics}
\label{app:fisher:results}

We evaluated the primary Fisher matrices at $T_{\rm eff}=3600$, 3800, and 4000~K and at $[Z]=0.0$ and $-0.5$, fixing $\log g=0.0$ and $\xi=3$~km~s$^{-1}$. Table~\ref{tab:fisher_results} gives the independent-information fractions, the gravity component of the weakest scaled eigenvector, and the condition numbers before and after adding the gravity prior.

\begin{table*}
\caption{Six-node Fisher identifiability diagnostics.}
\label{tab:fisher_results}
\centering
\small
\begin{tabular}{ccrrrrrrr}
\hline\hline
$T_{\rm eff}$ &
$[Z]$ &
$\eta_{T_{\rm eff}}$ &
$\eta_{\log g}$ &
$\eta_{\xi}$ &
$\eta_{[Z]}$ &
$\left|v_{{\rm weak},\log g}\right|$ &
$\kappa_{\rm spec}$ &
$\kappa_{\rm prior}$ \\
(K) &
(dex) &
(\%) &
(\%) &
(\%) &
(\%) &
&
&
\\
\hline
3600 &  0.0  & 13.6 & 3.7 & 31.3 &  8.8 & 0.964 & 1604 & 1323 \\
3800 &  0.0  & 17.7 & 4.0 & 32.7 &  7.9 & 0.959 & 1461 & 1239 \\
4000 &  0.0  & 38.3 & 6.4 & 34.8 &  8.0 & 0.958 & 1221 & 1064 \\
3600 & $-0.5$ & 17.2 & 4.5 & 31.3 & 10.2 & 0.963 & 1621 & 1210 \\
3800 & $-0.5$ & 21.9 & 4.5 & 32.2 &  7.8 & 0.957 & 1531 & 1181 \\
4000 & $-0.5$ & 36.2 & 7.0 & 34.2 &  7.9 & 0.961 & 1312 & 1041 \\
\hline
\end{tabular}
\tablefoot{$\eta_\alpha$ is the independent-information fraction defined in Sect.~\ref{app:fisher:eta}. The quantity $\left|v_{{\rm weak},\log g}\right|$ is the absolute coefficient of the scaled gravity coordinate in the weakest spectral Fisher eigenvector. The condition numbers $\kappa$ refer to the scaled spectral matrix and to the same matrix after addition of the gravity prior.}
\end{table*}

We additionally evaluated the solar-metallicity 3800~K model at $\log g=-0.25$ and $+0.5$. For $\log g=-0.25$, 0.0, and $+0.5$, the independent gravity fractions are 4.10, 3.98, and 4.20\%, the absolute gravity coefficients in the weakest mode are 0.958, 0.959, and 0.961, and the $\log g$--$[Z]$ correlations are 0.860, 0.849, and 0.821, respectively. The weak gravity identifiability is therefore not restricted to the central $\log g=0.0$ nodes.

To quantify the local propagation of a temperature offset, we partitioned the scaled Fisher matrix into the temperature coordinate and $\boldsymbol{u}=(\log g,\xi,[Z])$. The compensating displacement that minimises the local quadratic form for a fixed $\delta q_T$ is
\[
\delta\boldsymbol{q}_{u}
=
-(F_{uu})^{-1}F_{uT}\,\delta q_T.
\]
For $\delta T_{\rm eff}=+100$~K, the metallicity displacement is $+0.067$ to $+0.103$~dex for the spectral matrix and $+0.059$ to $+0.095$~dex after adding the 0.20-dex gravity prior to the nuisance block.

\label{app:fisher:wavelength}
Table~\ref{tab:fisher_wavelength_results} gives the raw and nuisance-orthogonal metallicity fractions accumulated by $1.22\,\mu{\rm m}$ at each node. It also lists the wavelengths by which 80 and 90\% of the nuisance-orthogonal information have accumulated.

\begin{table*}
\caption{Node-by-node cumulative metallicity-information diagnostics.}
\label{tab:fisher_wavelength_results}
\centering
\begin{tabular}{ccrrrr}
\hline\hline
$T_{\rm eff}$ &
$[Z]$ &
${\cal C}_{[Z]}^{\rm raw}(1.22\,\mu{\rm m})$ &
${\cal C}_{[Z]}^{\perp}(1.22\,\mu{\rm m})$ &
$\lambda_{80}^{\perp}$ &
$\lambda_{90}^{\perp}$ \\
(K) &
(dex) &
(\%) &
(\%) &
($\mu{\rm m}$) &
($\mu{\rm m}$) \\
\hline
3600 &  0.0  & 40.3 & 27.1 & 1.31250 & 1.31513 \\
3800 &  0.0  & 40.5 & 33.8 & 1.31228 & 1.31513 \\
4000 &  0.0  & 40.2 & 37.9 & 1.31206 & 1.31513 \\
3600 & $-0.5$ & 38.6 & 28.2 & 1.31250 & 1.31513 \\
3800 & $-0.5$ & 42.1 & 35.0 & 1.31228 & 1.31513 \\
4000 & $-0.5$ & 45.7 & 37.5 & 1.31184 & 1.31535 \\
\hline
\end{tabular}
\tablefoot{The cumulative fractions are normalised by the corresponding total information over the complete analysed interval. The quantities $\lambda_{80}^{\perp}$ and $\lambda_{90}^{\perp}$ are the wavelengths by which 80 and 90\% of the nuisance-orthogonal metallicity information have accumulated.}
\end{table*}

\section{Additional details of the spectral inference}
\label{app:inference_details}

This appendix gives the technical details of the gravity prior, continuum profiling, and regularised covariance introduced in Sect.~\ref{sec:inference}. It also records the secondary recovery diagnostics and qualifications for the mock experiments.

\subsection{Continuum implementations}
\label{app:continuum}

The emulator spectra were normalised relative to their synthetic continua. In the observations, however, line blanketing, molecular opacity, instrumental response, noise, and telluric residuals produce an empirical pseudo-continuum. We therefore profiled the relative continuum during each likelihood evaluation rather than applying a fixed normalisation beforehand.

For the adopted Gazak-like continuum treatment, we treated the three fitted intervals independently. At every trial atmospheric parameter vector, we ranked the valid model pixels according to $|\widetilde{m}_i-1|$, where $\widetilde{m}_i$ is the model flux before continuum correction, and selected the 15\% closest to the synthetic continuum as anchors after applying the wavelength masks. For each anchor pixel, we calculated
\[
q_i
=
\ln
\left(
\frac{\widetilde{m}_i}{f_{{\rm obs},i}}
\right).
\]
Within interval $j$, we fitted
\[
q_{{\rm fit},j}(x)
=
a_{0,j}+a_{1,j}x,
\]
where $x$ spans $[-1,1]$, and applied the multiplicative correction
\[
m_j(\lambda)
=
\frac{
\widetilde{m}_j(\lambda)
}{
\exp[q_{{\rm fit},j}(\lambda)]
}.
\]
The weighted fit was repeated at every likelihood evaluation, and the continuum coefficients were profiled rather than sampled as posterior parameters. Dynamic anchor selection allows the pseudo-continuum to follow changes in line blanketing with the atmospheric parameters and CNO mixture, while the constant and linear terms prevent the correction from reproducing narrower spectral structure.

The continuum experiment compared this method with the standard continuum procedure, which selected the highest points of the noisy observation, and an independent implementation of the exact model-guided algorithm of \citet{Gazak2014PhDT}. We generated ten noise realisations at $T_{\rm eff}=3800$~K, $\log g=0.0$, and $\xi=3$~km~s$^{-1}$ for each of $[Z]=0.0$ and $-0.5$. Each realisation was analysed with all three methods while the spectral grids, gravity prior, wavelength selection, noise array, and remaining inference configuration were held fixed, so the paired comparison isolated continuum placement. Figure~\ref{fig:continuum_mock} shows that the standard method produced a coherent negative metallicity displacement, whereas both model-guided implementations recovered the injected scale without a comparable systematic offset.

\subsection{Luminosity-based prior on surface gravity}
\label{app:gravity}

The extended $J$-band spectrum provides only limited independent information on surface gravity at the resolution considered here. We therefore introduced a luminosity-based prior whose centre depended on the observed $K$-band magnitude, visual extinction, adopted distance modulus, and the effective temperature sampled at each likelihood evaluation.

We used the flux-weighted gravity
\[
\log g_{\rm F}
=
\log g
-
4\log
\left(
\frac{T_{\rm eff}}{10^{4}\,{\rm K}}
\right),
\]
following \citet{Kudritzki2003}, together with the calibration
\[
\log g_{\rm F}
=
5.474
-
0.763
\log
\left(
\frac{L}{L_{\odot}}
\right).
\]
This calibration follows from the fundamental relations between gravity, mass, luminosity, and effective temperature combined with the approximate RSG relation $\log(M/M_{\odot})\simeq0.50-0.10M_{\rm bol}$. It provides a luminosity-based evolutionary expectation \citep[see][]{Levesque2005}.

At each likelihood evaluation, we calculated
\[
A_K=0.112A_V,
\qquad
BC_K
=
5.574
-
0.7589
\left(
\frac{T_{\rm eff}}{1000\,{\rm K}}
\right),
\]
where the bolometric correction is the Galactic RSG calibration of \citet{Levesque2005}. We then obtained
\[
M_{\rm bol}
=
m_K-\mu_0-A_K+BC_K,
\qquad
\log
\left(
\frac{L}{L_{\odot}}
\right)
=
-0.4
\left(
M_{\rm bol}-M_{{\rm bol},\odot}
\right),
\]
adopting $\mu_0=11.99$~mag for Per OB1 and $M_{{\rm bol},\odot}=4.75$~mag. The centre of the gravity prior at the sampled temperature was therefore
\[
\log g_{\rm prior}
=
4\log
\left(
\frac{T_{\rm eff}}{10^{4}\,{\rm K}}
\right)
+
5.474
-
0.763
\log
\left(
\frac{L}{L_{\odot}}
\right).
\]

We adopted the conditional Gaussian prior
\[
p
\left(
\log g
\mid
T_{\rm eff},m_K,A_V,\mu_0
\right)
\propto
\exp
\left[
-\frac{1}{2}
\left(
\frac{
\log g-\log g_{\rm prior}
}{
0.20
}
\right)^2
\right]
\]
within the atmospheric model-grid boundaries. We recalculated its centre at every sampled $T_{\rm eff}$ because both $BC_K$ and $\log g_{\rm prior}$ depend on temperature.

The width $\sigma_{\log g}=0.20$~dex encompasses uncertainties in distance, extinction, photometry, RSG variability, bolometric correction, and the approximate mass calibration. For reference, changing the distance modulus from 11.90 to 11.99~mag shifts the prior centre by only $-0.027$~dex, while an uncertainty of 0.35~mag corresponds to approximately 0.11~dex in gravity. For the controlled mocks, we instead centred the same prior directly on the injected value, $\log g=0.0$, preserving its regularising role without introducing artificial photometric uncertainty.

\subsection{Diagonal and regularised-covariance likelihoods}
\label{app:covariance}

For the residual vector $\boldsymbol{r}=\boldsymbol{f}_{\rm obs}-\boldsymbol{m}$, we used the Gaussian log-likelihood
\[
\ln{\cal L}
=
-\frac{1}{2}
\left[
\boldsymbol{r}^{\rm T}C^{-1}\boldsymbol{r}
+
\ln|C|
+
N_{\rm pix}\ln(2\pi)
\right].
\]
The diagonal covariance was
\[
C_{{\rm diag},ij}
=
\delta_{ij}\sigma_{{\rm diag},i}^{2},
\qquad
\sigma_{{\rm diag},i}^{2}
=
\sigma_{{\rm obs},i}^{2}
+
\sigma_{{\rm emul},i}^{2},
\]
so the emulator uncertainty was propagated together with the observational uncertainty. The Fisher diagnostics used the same independent-pixel form but the constant-$\mathrm{S/N}$ setup described in Sect.~\ref{sec:fisher}.

The regularised likelihood added a short-range covariance component,

\[
\mathbf{C}=\mathbf{C}_{\rm diag}+\mathbf{C}_{\rm cov}.
\]

As in our previous Cepheid analysis \citep[][]{Urbaneja2026b}, the covariance was constructed on
the observed wavelength grid, but the kernel distance was evaluated in
velocity space rather than in pixel or wavelength units. For each pair
of valid pixels \(i\) and \(j\), we defined the local velocity
separation as

\[
\Delta v_{ij}
 = c\,\frac{\lambda_j-\lambda_i}{(\lambda_i+\lambda_j)/2}.
\]

The correlated component was then

\[
C_{{\rm cov},ij}
 = \sigma_{{\rm diag},i}\sigma_{{\rm diag},j}s_{\rm cov}^{2}
   k\left(\frac{|\Delta v_{ij}|}{\ell_{v,ij}}\right),
\]

where the Matérn-3/2 kernel is

\[
k(d) = \left(1+\sqrt{3}\,d\right)\exp\left(-\sqrt{3}\,d\right).
\]

Here, \(s_{\rm cov}\) sets the correlated amplitude relative to the
diagonal uncertainty scale, and \(\ell_{v,ij}\) is the local velocity
scale obtained by converting the sampled pixel-space correlation
length \(\ell_{\rm cov}\) using the observed wavelength grid. Thus,
although \(\ell_{\rm cov}\) is sampled and reported in pixels, the
pairwise distance entering the kernel is evaluated in velocity space.
The correlated component was tapered beyond approximately one
correlation length using a Hann window, thereby restricting it to
local residual structure.

We used one global pair \((s_{\rm cov},\ell_{\rm cov})\) across the
three fitted intervals, and pixels excluded by the fixed masks
contributed to neither the residual vector nor the covariance matrix.

We sampled these hyperparameters together with the stellar and instrumental parameters, using the truncated priors
\[
\log s_{\rm cov}
\sim
{\cal N}(-1,0.3^2),
\qquad
-1.8
\leq
\log s_{\rm cov}
\leq
-0.2,
\]
and
\[
\ell_{\rm cov}
\sim
{\cal N}(2,1^2),
\qquad
1
\leq
\ell_{\rm cov}
\leq
6\ {\rm pixels}.
\]
The central values of these priors were informed by the
autocorrelation and amplitude of the residuals from the
diagonal-Gaussian analysis, which indicated structure on scales
of approximately two pixels and a characteristic relative
amplitude of $s_{\rm cov}\simeq0.1$. The truncation bounds keep
the covariance term confined to short-range residual structure,
while the log-determinant term and amplitude prior discourage
unsupported covariance amplitudes.

The covariance component redistributes coherent residual
structure between the atmospheric parameters and an explicit
residual process; it does not modify the synthetic spectrum.
Agreement between the diagonal and covariance analyses therefore
tests whether the metallicity depends strongly on forcing model
discrepancies into $T_{\rm eff}$, $\log g$, $\xi$, or $[Z]$.

For most Per~OB1 spectra and all three CNO grids, the posterior
hyperparameters cluster around $\log s_{\rm cov}\simeq-0.97$ and
$\ell_{\rm cov}\simeq2.1$ pixels. BD+56\,724 is the main
exception, with stronger and more extended residual structure,
as discussed in Sect.~5. We inspected the covariance posteriors
for contact with their imposed bounds. We therefore retained the
diagonal likelihood for the controlled mock experiments and used
the regularised covariance for the production and robustness
analyses of the observed spectra.

\subsection{Additional mock-recovery diagnostics}
\label{app:mock_diagnostics}

For the 18 matched-grid cases, the mean metallicity offset was
+0.029 dex, the RMS metallicity error was 0.035 dex, and the
maximum absolute offset was 0.06 dex. The RMS errors in
$T_{\rm eff}$, $\log g$, and $\xi$ were 8 K, 0.006 dex, and
0.16 km s$^{-1}$, respectively. A small common positive
metallicity offset near [Z]$_{\rm in}=-0.5$ appeared for all three
generating mixtures; because it was shared by the solar, mild,
and heavy cases, it is consistent with the paired noise
realisation and the local microturbulence--metallicity coupling
rather than with CNO composition. The recovery of all 18
injected metallicities within their nominal 68\% intervals is
therefore a descriptive consistency check, not a formal coverage
measurement based on repeated noise sampling.

Table~\ref{tab:cno_mismatch} records the mean
recovered-minus-injected metallicity for every mismatched pair of
generating and fitted CNO grids. The values were averaged over
the three input temperatures, while the matched-grid solutions
were excluded.

\begin{table}
\caption{Mean metallicity response to a fixed CNO-grid mismatch.}
\label{tab:cno_mismatch}
\centering
\begin{tabular}{llcc}
\hline\hline
True mixture &
Fitted grid &
\(\Delta[Z]_{0.0}\) & 
\(\Delta[Z]_{-0.5}\) \\
\hline
Solar & Mild  & $-0.050$ & $-0.063$ \\
Solar & Heavy & $-0.053$ & $-0.060$ \\
Mild  & Solar & $+0.047$ & $+0.047$ \\
Mild  & Heavy & $+0.003$ & $-0.007$ \\
Heavy & Solar & $ 0.000$ & $+0.027$ \\
Heavy & Mild  & $-0.040$ & $-0.017$ \\
\hline
\end{tabular}
\tablefoot{\(\Delta[Z]_{0.0}\) and \(\Delta[Z]_{-0.5}\) denote the
mean recovered-minus-injected metallicities, in dex, for input
metallicities of \(0.0\) and \(-0.5\)~dex, respectively, averaged
over the three input temperatures. The matched-grid solutions are
not included.}
\end{table}

Across the 18 generated mocks, the typical grid-dependent range
was approximately 0.05 dex in metallicity. The corresponding
ranges were 44 K in effective temperature and
0.13 km s$^{-1}$ in microturbulence. The maximum metallicity
range was 0.09 dex, while surface gravity changed by at most
0.02 dex because the adopted prior constrained the weakest
spectroscopic direction. Relative to the solar grid, the
processed grids generally returned slightly lower effective
temperatures and metallicities and slightly higher
microturbulent velocities. This redistribution explains why the
metallicity response did not follow a simple monotonic sequence
across the three mixtures.

We did not repeat the complete 54-fit matrix with the regularised
covariance or with additional noise seeds. The paired
grid-to-grid shifts varied smoothly across the six nodes, and the
shared realisations isolated the differential CNO trends from
common noise-induced offsets. Section~\ref{sec:perob1} instead
compares the two likelihoods for the observed spectra, where
structured residuals are present.

For the direct Fisher comparison discussed in
Sect.~\ref{sec:mocks_fisher_link}, we performed a separate
set of six matched-configuration mock analyses at the primary
Fisher nodes. In these tests, the calculation and Bayesian
inference were matched explicitly in wavelength selection,
continuum treatment, resolution, sampling, noise model,
parameter set, and gravity prior. 

Across the six nodes, the Fisher calculation reproduces the
marginal parameter uncertainties to within about 20\% for all four
parameters and captures the main parameter-degeneracy directions
seen in the Bayesian posteriors, including their variation from node
to node. Some weaker correlations differ more noticeably because the
full posterior explores a finite and non-linear region of parameter
space.

%As a check on ratios below unity, we examined the
%$T_{\rm eff}=3600$ K, $\log g=0.0$, $\xi=3$ km s$^{-1}$,
%$[Z]=-0.5$ node, which shows the largest metallicity discrepancy.
%Applying the model-grid boundaries to the corresponding Gaussian
%Fisher approximation reduces the expected posterior-to-Fisher
%width ratios substantially, reproducing most of the narrowing in
%$T_{\rm eff}$ and $\log g$ and part of that in $\xi$ and $[Z]$.
%The remaining difference is consistent with finite-volume
%non-linearity in the coupled spectral response.

The close agreement in both width and covariance structure
supports the use of the Fisher calculation as a local
information-content and identifiability diagnostic under the
controlled conditions of the matched mocks. It does not imply
that the observed-star posteriors should reproduce the same
curvature exactly, because those fits additionally include a
sampled resolving-power scale, structured residuals, a
luminosity-based gravity prior, and the regularised covariance
model described above.

\section{Individual results for Per OB1}
\label{app:perob1}

\subsection{Individual posterior results}

Table~\ref{tab:perob1_individual} gives the individual posterior results for the preferred mild-CNO grid with the regularised global-covariance likelihood. We list the two observations of BD+59~372 separately, so the table contains 21 spectral analyses representing 20 unique stars. Section~\ref{sec:perob1_results} counted this object only once when calculating the association statistics.

We report the posterior median and the corresponding 16th and 84th percentile offsets for the four atmospheric parameters, the resolving-power scale factor, and the two covariance hyperparameters. The CNO mixture was fixed by the selected grid and was not inferred independently for each star. The surface gravities include the luminosity-based prior and are therefore not purely spectroscopic determinations, while the covariance hyperparameters describe the adopted residual model rather than stellar properties.

For most spectra, the covariance hyperparameters clustered around $\log s_{\rm cov}\simeq-0.97$ and $\ell_{\rm cov}\simeq2.1$ pixels, with no appreciable dependence on the CNO grid. BD+56~724 was the clear exception, with representative values near $-0.24$ and 3.8 pixels, respectively, indicating stronger and more extended structured residuals. Excluding this object from the preferred configuration changed the association mean from $-0.095$ to $-0.100$~dex and the dispersion from 0.089 to 0.088~dex, confirming that it does not control the association result.

\begin{table*}
\caption{Individual Per OB1 results from the preferred mild-CNO, regularised-covariance analysis.}
\label{tab:perob1_individual}
\centering
\scriptsize
\setlength{\tabcolsep}{3pt}
\renewcommand{\arraystretch}{1.08}
\begin{tabular}{lccccccc}
\hline\hline
Star &
$T_{\rm eff}$ &
$\log g$ &
$\xi$ &
$[Z]$ &
$s_R$ &
$\log s_{\rm cov}$ &
$\ell_{\rm cov}$ \\
 &
(K) &
(dex) &
(km\,s$^{-1}$) &
(dex) &
 &
 &
(pixels) \\
\hline
BD+59~372
& $3838_{-179}^{+173}$
& $ 0.62_{-0.12}^{+0.18}$
& $3.0_{-0.7}^{+0.8}$
& $-0.13_{-0.27}^{+0.19}$
& $1.04_{-0.14}^{+0.11}$
& $-0.99_{-0.17}^{+0.15}$
& $2.06_{-0.94}^{+0.84}$ \\

BD+59~372 (2)
& $3844_{-183}^{+164}$
& $ 0.61_{-0.16}^{+0.15}$
& $3.0_{-0.8}^{+0.8}$
& $-0.17_{-0.24}^{+0.24}$
& $1.03_{-0.16}^{+0.11}$
& $-0.99_{-0.16}^{+0.15}$
& $2.08_{-1.01}^{+0.84}$ \\

BD+56~595
& $3779_{-149}^{+143}$
& $ 0.27_{-0.12}^{+0.17}$
& $3.7_{-0.7}^{+0.8}$
& $-0.13_{-0.20}^{+0.21}$
& $1.00_{-0.12}^{+0.10}$
& $-0.99_{-0.15}^{+0.14}$
& $2.13_{-1.09}^{+0.71}$ \\

HD~14404
& $3763_{-180}^{+141}$
& $ 0.14_{-0.16}^{+0.14}$
& $3.7_{-0.9}^{+0.6}$
& $-0.12_{-0.18}^{+0.19}$
& $1.03_{-0.11}^{+0.10}$
& $-0.97_{-0.17}^{+0.16}$
& $2.14_{-0.98}^{+0.74}$ \\

HD~14826
& $3671_{-151}^{+125}$
& $ 0.06_{-0.16}^{+0.15}$
& $3.3_{-0.7}^{+0.7}$
& $-0.06_{-0.20}^{+0.20}$
& $1.02_{-0.11}^{+0.11}$
& $-0.98_{-0.17}^{+0.14}$
& $2.18_{-0.98}^{+0.77}$ \\

HD~236979
& $3934_{-162}^{+132}$
& $ 0.05_{-0.14}^{+0.16}$
& $3.0_{-0.6}^{+0.6}$
& $ 0.12_{-0.15}^{+0.18}$
& $1.01_{-0.09}^{+0.10}$
& $-0.97_{-0.17}^{+0.15}$
& $2.14_{-0.89}^{+0.91}$ \\

HD~13136
& $3722_{-134}^{+144}$
& $-0.02_{-0.17}^{+0.13}$
& $3.8_{-0.7}^{+0.7}$
& $-0.16_{-0.20}^{+0.18}$
& $1.03_{-0.11}^{+0.09}$
& $-0.98_{-0.15}^{+0.17}$
& $2.16_{-0.97}^{+0.81}$ \\

HD~14270
& $3683_{-160}^{+145}$
& $ 0.03_{-0.14}^{+0.16}$
& $3.2_{-0.8}^{+0.7}$
& $-0.13_{-0.19}^{+0.22}$
& $1.02_{-0.13}^{+0.10}$
& $-0.97_{-0.17}^{+0.13}$
& $2.09_{-0.90}^{+0.84}$ \\

BD+56~724
& $3558_{-146}^{+117}$
& $-0.12_{-0.15}^{+0.12}$
& $3.8_{-0.8}^{+0.9}$
& $ 0.01_{-0.19}^{+0.22}$
& $1.00_{-0.13}^{+0.10}$
& $-0.24_{-0.01}^{+0.04}$
& $3.82_{-0.55}^{+0.73}$ \\

HD~14469
& $3579_{-105}^{+109}$
& $-0.14_{-0.15}^{+0.15}$
& $3.9_{-0.8}^{+0.6}$
& $-0.12_{-0.17}^{+0.17}$
& $1.04_{-0.12}^{+0.08}$
& $-0.95_{-0.19}^{+0.15}$
& $2.16_{-0.93}^{+0.86}$ \\

HD~14488
& $3545_{-141}^{+99}$
& $-0.12_{-0.17}^{+0.12}$
& $3.3_{-0.6}^{+0.7}$
& $-0.16_{-0.18}^{+0.20}$
& $0.99_{-0.10}^{+0.12}$
& $-0.99_{-0.16}^{+0.15}$
& $2.13_{-0.98}^{+0.85}$ \\

BD+57~530A
& $3706_{-167}^{+182}$
& $ 0.35_{-0.15}^{+0.15}$
& $3.3_{-0.8}^{+0.7}$
& $-0.19_{-0.22}^{+0.22}$
& $0.98_{-0.12}^{+0.11}$
& $-0.98_{-0.18}^{+0.14}$
& $2.08_{-1.12}^{+0.66}$ \\

BD+57~647
& $3769_{-148}^{+128}$
& $ 0.08_{-0.16}^{+0.13}$
& $3.5_{-0.7}^{+0.6}$
& $ 0.04_{-0.17}^{+0.18}$
& $1.00_{-0.10}^{+0.09}$
& $-0.96_{-0.15}^{+0.17}$
& $2.12_{-0.90}^{+0.89}$ \\

BD+58~501
& $3695_{-142}^{+120}$
& $ 0.03_{-0.16}^{+0.14}$
& $3.4_{-0.7}^{+0.7}$
& $-0.06_{-0.20}^{+0.19}$
& $1.03_{-0.11}^{+0.10}$
& $-0.97_{-0.16}^{+0.18}$
& $2.15_{-1.03}^{+0.76}$ \\

BD+56~512
& $3697_{-141}^{+112}$
& $ 0.12_{-0.14}^{+0.17}$
& $4.1_{-0.7}^{+0.7}$
& $-0.06_{-0.17}^{+0.17}$
& $1.01_{-0.10}^{+0.09}$
& $-0.96_{-0.17}^{+0.16}$
& $2.23_{-1.00}^{+0.79}$ \\

HD~13658
& $3839_{-168}^{+148}$
& $ 0.45_{-0.16}^{+0.14}$
& $3.2_{-0.8}^{+0.6}$
& $-0.11_{-0.23}^{+0.18}$
& $1.04_{-0.12}^{+0.11}$
& $-0.99_{-0.16}^{+0.14}$
& $2.07_{-0.93}^{+0.84}$ \\

HD~14242
& $3874_{-137}^{+146}$
& $ 0.21_{-0.14}^{+0.16}$
& $3.7_{-0.8}^{+0.6}$
& $-0.05_{-0.20}^{+0.16}$
& $1.00_{-0.09}^{+0.09}$
& $-0.96_{-0.16}^{+0.16}$
& $2.16_{-0.96}^{+0.84}$ \\

HD~14330
& $3811_{-152}^{+148}$
& $ 0.20_{-0.16}^{+0.13}$
& $3.4_{-0.6}^{+0.7}$
& $-0.01_{-0.21}^{+0.16}$
& $1.02_{-0.12}^{+0.10}$
& $-0.99_{-0.17}^{+0.14}$
& $2.11_{-0.94}^{+0.84}$ \\

HD~236915
& $3646_{-175}^{+136}$
& $ 0.20_{-0.15}^{+0.15}$
& $3.6_{-0.8}^{+0.7}$
& $-0.25_{-0.22}^{+0.22}$
& $0.98_{-0.13}^{+0.10}$
& $-0.97_{-0.17}^{+0.14}$
& $2.11_{-0.94}^{+0.86}$ \\

HD~236947
& $3682_{-169}^{+139}$
& $ 0.24_{-0.13}^{+0.17}$
& $3.2_{-0.8}^{+0.6}$
& $-0.21_{-0.21}^{+0.23}$
& $1.04_{-0.13}^{+0.11}$
& $-0.98_{-0.16}^{+0.15}$
& $2.12_{-0.83}^{+1.05}$ \\

HD~14580
& $3870_{-186}^{+136}$
& $ 0.39_{-0.14}^{+0.16}$
& $3.2_{-0.8}^{+0.6}$
& $-0.09_{-0.21}^{+0.18}$
& $1.01_{-0.13}^{+0.10}$
& $-1.00_{-0.15}^{+0.15}$
& $2.07_{-1.01}^{+0.83}$ \\
\hline
\end{tabular}
\tablefoot{The entries give the posterior median and the offsets to the 16th and 84th percentiles. The results use the mildly processed CNO grid, Gazak-like continuum treatment, luminosity-based gravity prior, and regularised global covariance. The resolving-power scale factor is defined by $R_{\rm eff}=s_RR_{\rm nom}$, with $R_{\rm nom}=2000$. The covariance amplitude \(s_{\rm cov}\) is dimensionless relative to
the diagonal pixel uncertainties. The correlation length
\(\ell_{\rm cov}\) is sampled and reported in pixels and is converted
internally to the local velocity scale used to evaluate the covariance
kernel, as described in Appendix~\ref{app:covariance}.}
\end{table*}

\begin{sidewaysfigure*}
\centering
\includegraphics{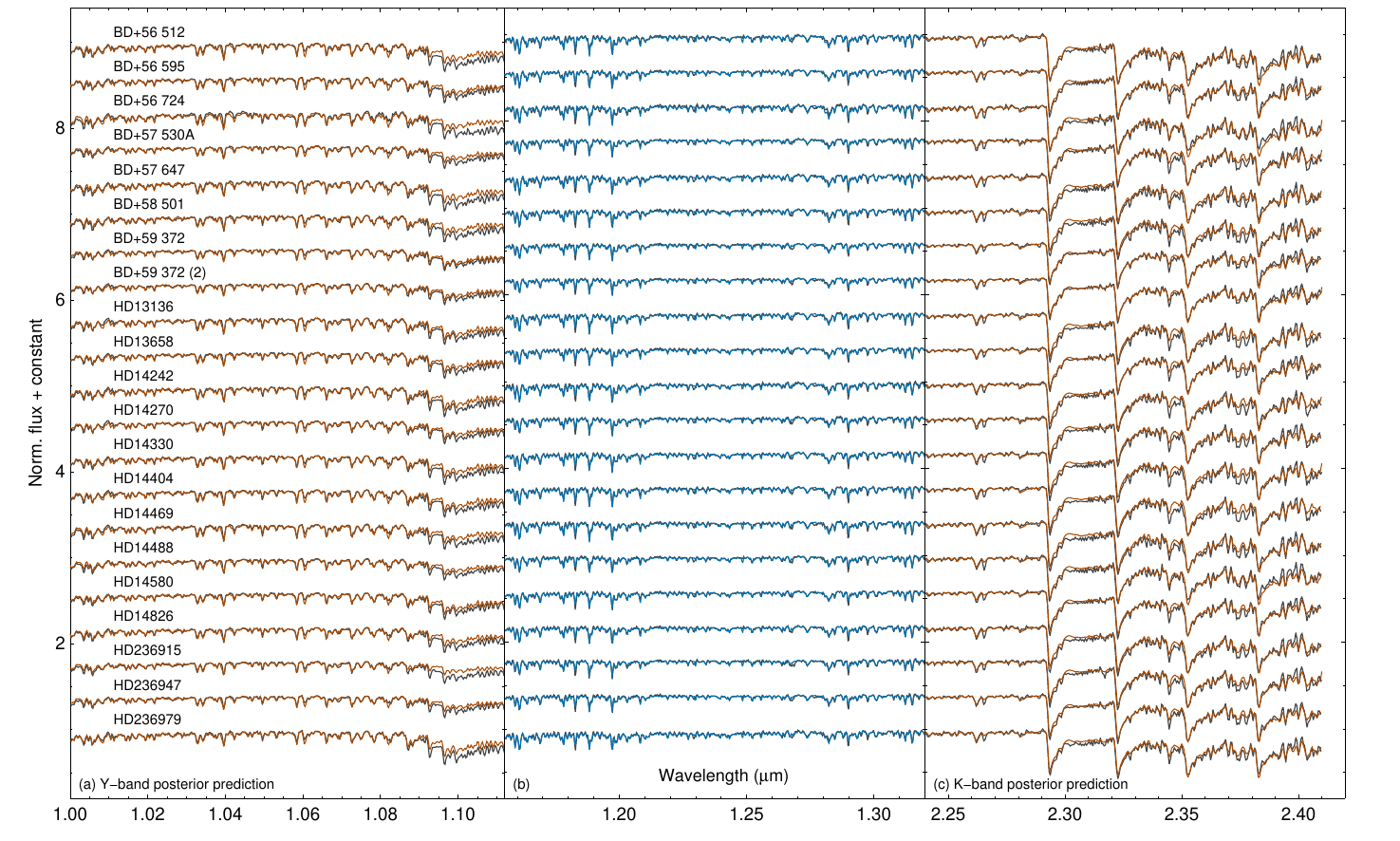}
\caption{Global comparison between the observed spectra and the preferred model spectra for the Per OB1 sample. Black and colour lines show the observations and models, respectively, in the $Y$, $J$, and $K$ bands. Only the retained pixels in the extended $J$-band interval contributed to the inference; the $Y$- and $K$-band models are posterior predictions. The spectra are shifted vertically and ordered by identifier, with both observations of BD+59~372 shown separately.}
\label{fig:global_fits}
\end{sidewaysfigure*}

Figure~\ref{fig:global_fits} compares the observed spectra with the preferred posterior models for the complete Per~OB1 sample. Only the extended \(J\)-band interval contributed to the inference, while the \(Y\)- and \(K\)-band models provide posterior predictions outside the fitted wavelength range. We note that Fig.~\ref{fig:global_fits} shows a systematic residual near
\(1.1\,\mu{\rm m}\), outside the fitted wavelength range, where the posterior
predictions underestimate the depth of the observed CN absorption. The
mismatch is present for essentially every star in the sample and is
not sensitive to the adopted CNO grid, suggesting a shortcoming of the
model framework in this band rather than an abundance-pattern effect
specific to Per~OB1. A qualitatively similar CN deficit near
\(1.1\,\mu{\rm m}\) was reported by \citet[][]{Lancon2007} for PHOENIX models of
red supergiants and giants, where it persisted even after adopting
RSG-specific surface abundances. Since this feature falls outside the extended $J$-band interval used in
the inference, it does not affect the fitted parameters, but it points to
a broader limitation in modelling the CN-dominated region around
\(1.1\,\mu{\rm m}\) that appears across independent one-dimensional model
frameworks and may be worth revisiting in future
MARCS/TURBOSPECTRUM developments.

\subsection{Gravity-prior sensitivity}

We repeated the preferred mild-CNO covariance analysis with the broad fixed priors $\log g\sim\mathcal{N}(0.0,0.5^2)$ and $\log g\sim\mathcal{N}(0.2,0.5^2)$. Relative to the luminosity-based prior, they changed the association mean by $-0.028$ and $-0.006$~dex, respectively, while the star-to-star dispersion remained between 0.089 and 0.106~dex. The typical posterior gravity uncertainty approximately doubled, whereas microturbulence changed little.

Across the 40 shifts from 20 unique stars and two alternative priors, the Pearson correlation coefficients were $r=0.98$ between $\Delta T_{\rm eff}$ and $\Delta\log g$ and $r=0.97$ between $\Delta[Z]$ and $\Delta\log g$. Lower inferred gravities were accompanied by lower effective temperatures and metallicities, following the coupled Fisher directions. The luminosity-based prior therefore constrains the location along the weakest mode without setting the association metallicity scale.

\begin{table*}[!h]
\caption{Atmospheric parameters from the repeatability and external
comparison spectra.}
\label{tab:repeatability_comparisons}
\centering
\scriptsize
\setlength{\tabcolsep}{4pt}
\begin{tabular}{llcccc}
\hline\hline
Target &
Spectrum or analysis &
\(T_{\rm eff}\) &
\(\log g\) &
\(\xi\) &
Abundance \\
&
&
(K) &
(dex) &
(\({\rm km\,s^{-1}}\)) &
(dex) \\
\hline
BD+59~372
 & Programme SpeX (1)
 & \(3838_{-179}^{+173}\)
 & \(0.62_{-0.12}^{+0.18}\)
 & \(3.0_{-0.7}^{+0.8}\)
 & \(-0.13_{-0.27}^{+0.19}\) \\
BD+59~372
 & Programme SpeX (2)
 & \(3844_{-183}^{+164}\)
 & \(0.61_{-0.16}^{+0.15}\)
 & \(3.0_{-0.8}^{+0.8}\)
 & \(-0.17_{-0.24}^{+0.24}\) \\
\hline
HD~14404
 & Programme SpeX
 & \(3763_{-180}^{+141}\)
 & \(0.14_{-0.16}^{+0.14}\)
 & \(3.7_{-0.9}^{+0.6}\)
 & \(-0.12_{-0.18}^{+0.19}\) \\
HD~14404
 & IRTF-library SpeX
 & \(3785_{-154}^{+134}\)
 & \(0.15_{-0.14}^{+0.16}\)
 & \(3.9_{-0.8}^{+0.7}\)
 & \(-0.20_{-0.21}^{+0.16}\) \\
%\addlinespace
\hline
HD~14469
 & Programme SpeX
 & \(3579_{-105}^{+109}\)
 & \(-0.14_{-0.15}^{+0.15}\)
 & \(3.9_{-0.8}^{+0.6}\)
 & \(-0.12_{-0.17}^{+0.17}\) \\
HD~14469
 & IRTF-library SpeX
 & \(3560_{-108}^{+127}\)
 & \(-0.15_{-0.13}^{+0.16}\)
 & \(3.7_{-0.7}^{+0.6}\)
 & \(-0.13_{-0.18}^{+0.17}\) \\
%\addlinespace
\hline
HD~14488
 & Programme SpeX
 & \(3545_{-141}^{+99}\)
 & \(-0.12_{-0.17}^{+0.12}\)
 & \(3.3_{-0.6}^{+0.7}\)
 & \(-0.16_{-0.18}^{+0.20}\) \\
HD~14488
 & IRTF-library SpeX
 & \(3537_{-129}^{+100}\)
 & \(-0.12_{-0.16}^{+0.13}\)
 & \(3.0_{-0.7}^{+0.6}\)
 & \(-0.18_{-0.21}^{+0.21}\) \\
\hline
V424~Lac
 & IRTF-library SpeX, this work
 & \(3792_{-193}^{+174}\)
 & \(0.55_{-0.16}^{+0.14}\)
 & \(2.8_{-0.7}^{+0.7}\)
 & \(-0.04_{-0.24}^{+0.21}\) \\
V424~Lac
 & MAGIS, this work
 & \(3784_{-103}^{+98}\)
 & \(0.52_{-0.16}^{+0.13}\)
 & \(2.8_{-0.4}^{+0.3}\)
 & \(+0.04_{-0.15}^{+0.16}\) \\
V424~Lac
 & MAGIS, T25
 & \(3767_{-48}^{+48}\)
 & \(0.49_{-0.13}^{+0.12}\)
 & \(1.98_{-0.11}^{+0.13}\)
 & \(-0.039_{-0.035}^{+0.039}\) \\
%\addlinespace
\hline
Betelgeuse
 & IRTF-library SpeX, this work
 & \(3599_{-121}^{+138}\)
 & \(-0.04_{-0.16}^{+0.13}\)
 & \(3.8_{-0.7}^{+0.8}\)
 & \(-0.09_{-0.17}^{+0.22}\) \\
Betelgeuse
 & MAGIS, this work
 & \(3757_{-83}^{+97}\)
 & \(-0.03_{-0.14}^{+0.15}\)
 & \(3.6_{-0.4}^{+0.5}\)
 & \(-0.12_{-0.12}^{+0.16}\) \\
Betelgeuse
 & MAGIS, T25
 & \(3633_{-37}^{+37}\)
 & \(-0.06_{-0.15}^{+0.14}\)
 & \(2.19_{-0.17}^{+0.16}\)
 & \(-0.111_{-0.060}^{+0.076}\) \\
\hline
\end{tabular}
\tablefoot{For the analyses performed in this work, the entries give
posterior medians and offsets to the 16th and 84th percentiles. The
rows labelled T25 reproduce the central values and uncertainties
reported by \citet{Taniguchi2025}. All results obtained in this work
use the mildly processed CNO grid, Gazak-like continuum treatment,
luminosity-based gravity prior, and regularised covariance likelihood.
The abundance column gives the global \([Z]\) coordinate for our
analyses and \([\mathrm{Fe/H}]\) for the independent
MAGIS analysis of \citet[][T25]{Taniguchi2025}. Programme SpeX denotes the
Per~OB1 observations, while IRTF-library SpeX denotes spectra from
the IRTF Spectral Library \citep{Rayner2009}.}
\end{table*}

\subsection{Repeatability and external comparisons}

Table~\ref{tab:repeatability_comparisons} gives the atmospheric
parameters from the programme repeat and the external comparison
spectra. In the preferred analysis, the two observations of
BD+59~372 gave \([Z]=-0.13\) and \(-0.17\)~dex, while the other
atmospheric parameters were essentially unchanged. Across all six
combinations of CNO grid and likelihood, their absolute metallicity
difference ranged from 0.00 to 0.10~dex and remained smaller than
the individual 68\% posterior intervals.

Three programme targets, HD~14404, HD~14469, and HD~14488, have
independent spectra in the IRTF Spectral Library
\citep{Rayner2009}. In the matched mild-CNO covariance analysis,
the archival-minus-programme metallicity differences were
\(-0.08\), \(-0.01\), and \(-0.02\)~dex, respectively, giving a
mean of \(-0.04\)~dex and an RMS of 0.05~dex. The corresponding
absolute differences did not exceed 22~K in \(T_{\rm eff}\),
0.01~dex in \(\log g\), or \(0.3\,{\rm km\,s^{-1}}\) in \(\xi\).

As a further external check, we directly fitted the native-resolution
MAGIS spectra of V424~Lac and Betelgeuse at
\(R\simeq28\,000\) \citep{Taniguchi2025}. Relative to their IRTF
SpeX spectra, the metallicity differences were \(-0.08\) and
\(+0.03\)~dex, respectively. The remaining parameters agree closely
for V424~Lac; for Betelgeuse, \(\log g\) and \(\xi\) agree, while
\(T_{\rm eff}\) differs by approximately 160~K. The SpeX results are also 
broadly consistent with the independent
MAGIS analysis of \citet{Taniguchi2025}. For V424~Lac and
Betelgeuse, respectively, the temperature differences are \(+25\)
and \(-34\)~K, the gravity differences are \(+0.06\) and
\(+0.02\)~dex, and the differences between our global \([Z]\)
coordinate and their \([\mathrm{Fe/H}]\) values are approximately
0.00 and \(+0.02\)~dex. The SpeX microturbulent velocities are
higher by approximately 0.8 and \(1.6\,{\rm km\,s^{-1}}\).

These comparisons differ in instrument, observing epoch, resolution,
and reduction, and are too few to define an external abundance
calibration. Together, however, the programme repeat, IRTF-library
pairs, and MAGIS comparisons reveal no metallicity displacement
larger than the typical individual 68\% metallicity uncertainty of about
0.2 dex.

\end{appendix}

\end{document}